\documentclass[12pt]{iopart}

\usepackage[utf8]{inputenc}
\usepackage{amssymb}
\expandafter\let\csname equation*\endcsname\relax
\expandafter\let\csname endequation*\endcsname\relax

\usepackage{amsmath}
\usepackage{bbm}
\usepackage{bm}
\usepackage[]{todonotes}
\usepackage{mathtools}
\usepackage{tikz}
\usepackage{tikz-feynman}
\usepackage{xcolor}
\usepackage{cite}

\newif\ifediting

\editingtrue

\ifediting 
\newcommand\docNote[1]{
 \todo[backgroundcolor=blue!20!white,fancyline,
 bordercolor=white]{DOC: #1}}

\newcommand\dakNote[1]{
 \todo[backgroundcolor=blue!30!white,fancyline,
 bordercolor=white]{DAK: #1}}

\newcommand\rmNote[1]{
 \todo[backgroundcolor=blue!10!white,fancyline,
 bordercolor=white]{RM: #1}}
\else 
\newcommand\docNote[1]{}

\newcommand\dakNote[1]{}

\newcommand\rmNote[1]{}
\fi

\renewcommand{\[}{\begin{equation}\begin{aligned}}
\renewcommand{\]}{\end{aligned}\end{equation}}

\newcommand{\hel}{\eta} % Helicity label
\newcommand{\kb}{\bar k}
\newcommand{\qb}{\bar q}

\renewcommand{\d}{\mathrm{d}}
\newcommand{\dd}{\hat{\mathrm{d}}}
\newcommand{\del}{\hat{\delta}}
\newcommand{\dPhi}{\d \Phi}
\newcommand{\Del}{\hat{\delta}_\Phi}

\newcommand{\ket}[1]{| #1 \rangle}
\newcommand{\bra}[1]{\langle #1 |}

\newcommand{\Oop}{\mathbb{O}}
\newcommand{\Fop}{\mathbbm{F}}

\newcommand{\Rop}{\mathbbm{R}}

\def\Lexp{\biggl\langle\!\!\!\biggl\langle}
\def\Rexp{\biggr\rangle\!\!\!\biggr\rangle}

\renewcommand{\Re}{\operatorname{Re}}

\DeclareMathOperator{\sign}{sign}

\renewcommand{\v}[1]{\mathbf{#1}}

\def\creation#1{a^\dagger_{#1}}
\def\annihilation#1{a^{\vphantom{\dagger}}_{#1}}

\def\braa#1{% \vspace*{4mm} 
        \left\langle\smash{#1}{\vphantom1}\right|}

\def\keta#1{%
        \left|\smash{#1}{\vphantom1}\right\rangle}

\def\timecomponentlabel{0}

\def\Del{\del_\Phi}
\def\wn{\bar}
\def\dd{\hat d}
\def\del{\hat \delta}
\def\delp{\hat \delta^{(+)}}
\def\vpt{{\vphantom{2}}}
\renewcommand{\v}[1]{\boldsymbol{{#1}}}

\renewcommand{\Re}{\operatorname{Re}}
\def\vpp{{\vphantom{\prime}}} % to match subscript vertical positions
\def\Eqn#1{Eq.~(\ref{#1})}
\def\eqn#1{eq.~(\ref{#1})}
\def\Ampl{\mathcal{A}}
\def\AmplB{\mathcal{\overline {\kern-4pt A}}}
\def\sumint#1{\mathclap{\displaystyle\int}\mathclap{\displaystyle\sum_{#1}}}
\def\lsumint#1{\hspace*{1.5mm}{}\mathclap{\textstyle\int}\hspace*{1.5mm}\mathclap{\textstyle\sum_{#1}}\hspace*{2.5mm}}
\def\operator{\mathbb}
\def\ImpA{I_{(1)}^\mu}
\def\ImpB{I_{(2)}^\mu}
\def\ImpAcl{I_{(1),\class}^\mu}

\newcommand{\ImpAclsup}[1]{I_{(1),\class}^{\mu,#1}}

\definecolor{allOrderBlue}{rgb}{0.4,0.5,1}
\definecolor{patternBlue}{rgb}{0,0,1}
\definecolor{photonRed}{rgb}{1,0.2,0.2}
\def\xfer{w}

\newcommand{\DeltaPlo}{\Delta p_{1}^{\mu, (0)}}
\def\class{\textrm{cl}}
\def\sect#1{sect.~\ref{#1}}

\begin{document}
\bibliographystyle{iopart-num}

\begin{flushright}
	SAGEX-22-15\\
	QMUL-PH-22-08
\end{flushright}

\title[Classical Gravity From Scattering Amplitudes]{The SAGEX Review on Scattering Amplitudes \\ Chapter 14: Classical Gravity from Scattering Amplitudes}

\author{David~A.~Kosower$^{1}$}
\address{$1$ Institut de Physique Th\'eorique, CEA, CNRS, Universit\'e Paris--Saclay,\\ F--91191 Gif-sur-Yvette cedex, France}
\ead{David.Kosower@ipht.fr}

\author{Ricardo Monteiro$^{2}$}
\address{$2$ Centre for Theoretical Physics, Department of Physics and Astronomy, \\ Queen Mary University of London, E1 4NS, England, UK}
\ead{ricardo.monteiro@qmul.ac.uk}

\author{Donal O'Connell$^{3}$}
\address{$3$ Higgs Centre for Theoretical Physics, School of Physics and Astronomy, \\ The University of Edinburgh, Edinburgh EH9~3JZ, Scotland, UK}
\ead{donal@ed.ac.uk}

\begin{abstract}
Scattering amplitudes have their origin in quantum field theory, 
but have wide-ranging applications extending to classical physics.
We review a formalism to connect certain classical observables 
to scattering amplitudes. 
An advantage of this formalism is that it enables us to study implications
of the double copy in classical gravity. 
We discuss examples of observables including the total change of a particle's momentum, 
and the gravitational waveform, during a scattering encounter.
The double copy also allows direct access to classical solutions in gravity. 
We review this classical double copy starting from its linearised level, where it originates in the double copy of three-point amplitudes. 
The classical double copy extends elegantly to exact solutions, making a connection between scattering amplitudes and the geometric formulation of General Relativity.
\end{abstract}

%\maketitle
\newpage

\section{Introduction}
\label{sec:intro}

At the end of a century of experimental and theoretical work, we
know that a theory of Nature must respect both the laws of
quantum mechanics (QM) and the framework of special relativity (SR).
We have tested the forces of particle physics in detail, and have seen
beautiful agreement with both.
Gravity, ironically the one facet of fundamental
physics whose rudiments
were known to the ancients, stands apart.
The spectacular observational tests of Einstein gravity 
have been restricted to classical aspects.
Observing genuinely quantum aspects of gravity presumably requires
probing extremely short distances experimentally, which is very difficult.

Theorists are not so easily corralled, however.  The late
Steven Weinberg taught us how to control theories, including
gravity, even in the ignorance of their short-distance behaviour.
Furthermore, the underlying quantum nature of the rest of physical
theory ensures that we can choose to see any classical measurement
as a limit of a quantum one.

We can thus choose to compute 
the results of any measurement in classical gravity 
using quantum-mechanical methods.  What
is more surprising is that it can be \textit{useful\/} to do
so.  This surprise, and a particular framework for doing so,
are the motivation for this review.

The surprise has its roots in a relation first noticed in string theory~\cite{Kawai:1985xq}. 
In the field-theory limit of string theory,
this relation connects amplitudes in gravity theories to
`squares' of amplitudes in Yang--Mills (YM) theories~\cite{Bern:1991aq,Cachazo:2013hca}.  
A different way of formulating such `double copy' relations
presents them as a consequence of a duality between
YM colour and kinematic degrees of freedom~\cite{Bern:2008qj,Bern:2010ue,Bern:2010yg}.  This
duality can be implemented at a diagrammatic level, 
providing a sharp tool useful for loop calculations as well as tree-level ones.
In particular, the duality has direct relevance for classical gravitational radiation~\cite{Luna:2016due,Goldberger:2016iau,Shen:2018ebu,Luna:2017dtq}.

In this review, we start by presenting the connection of scattering amplitudes to
observables using an on-shell formalism developed by Maybee and two of the authors (KMOC)~\cite{Kosower:2018adc}.
It starts with quantum observables, expresses them in terms
of scattering amplitudes, and then offers evaluations of
the corresponding classical quantities.  In this setting,
four-point amplitudes determine quantities such as
the scattering angle, with five-point amplitudes determining
radiation.
The expression for the waveform in terms of amplitudes~\cite{Cristofoli:2021vyo}
establishes the double copy's utility for computing
the waveform to all orders in perturbation theory.
Other researchers have applied and extended these ideas in a number of important 
directions~\cite{Maybee:2019jus,Cristofoli:2020hnk,Manu:2020zxl,%
delaCruz:2020bbn,delaCruz:2020cpc,Herrmann:2021lqe,Herrmann:2021tct,Aoude:2021oqj,%
delaCruz:2021gjp,Bautista:2021llr,Cristofoli:2021jas,Bern:2021xze,Britto:2021pud,%
Manohar:2022dea}.

We further review how the double copy applies directly to classical solutions
in YM theory and gravity.
The starting point is the double copy of
three-point scattering amplitudes, studied
in a spacetime with $(2,2)$ signature instead of conventional
Minkowski space. These amplitudes determine the 
linearised YM field strength or the space-time curvature of classical gravitational solutions. The 
double copy relating the amplitudes then translates into a double 
copy relating the classical solutions. Remarkably, for the most 
important exact solutions of General Relativity (GR) --- Schwarzschild 
and Kerr --- this double-copy relation extends naturally beyond 
the linearised level to an exact statement.  
The Kerr--Schild~\cite{Monteiro:2014cda} and Weyl~\cite{Luna:2018dpt} double copies are exact formulations of
the classical relation. 
These advances show that insights from scattering amplitudes into classical gravity extend beyond the perturbative realm, and have a deep connection to the geometric formulation of GR. We give an overview of the achievements and limitations of the classical double copy, including some open questions.

Researchers have developed many other approaches to conservative and radiative processes in 
the gravitational two-body problem.
These include direct solutions of the GR equations, either perturbative~\cite{Kovacs:1978eu,Buonanno:1998gg,Damour:2000we,Damour:2001bu,Blanchet:2013haa,Barack:2018yvs} or numerical~\cite{Pretorius:2005gq,Bishop:2016lgv}; effective field theory~\cite{Goldberger:2004jt,Goldberger:2009qd,Foffa:2011ub,Foffa:2012rn,Levi:2015msa,Porto:2016pyg,Foffa:2016rgu,Levi:2018nxp,Endlich:2016jgc} or worldline~\cite{Kalin:2020mvi,Mogull:2020sak,Jakobsen:2021smu,Mougiakakos:2021ckm,Jakobsen:2021lvp,Jakobsen:2021zvh,Jakobsen:2022fcj} methods; and
use of the eikonal approximation~\cite{Torgerson:1966zz,Abarbanel:1969ek,Levy:1969cr,Amati:1987wq,%
tHooft:1987vrq,Muzinich:1987in,Amati:1987uf,Kabat:1992tb,Saotome:2012vy,Melville:2013qca,Akhoury:2013yua,%
Luna:2016idw,Ciafaloni:2018uwe,DiVecchia:2019myk,KoemansCollado:2019ggb,Bern:2020gjj,Parra-Martinez:2020dzs,%
DiVecchia:2020ymx,DiVecchia:2021ndb,DiVecchia:2021bdo,Adamo:2021rfq}.
Approaches based on the modern amplitudes programme include reconstructing the classical potential from four-point
amplitudes~\cite{Donoghue:1993eb,Donoghue:1994dn,Donoghue:2001qc,Bjerrum-Bohr:2002fji,Bjerrum-Bohr:2002gqz,%
Neill:2013wsa,Bjerrum-Bohr:2013bxa,Cachazo:2017jef,Bjerrum-Bohr:2018xdl,Cristofoli:2019neg,%
Bjerrum-Bohr:2019kec,Cristofoli:2020uzm,Bjerrum-Bohr:2021vuf,Bjerrum-Bohr:2021din}, matching to effective field theory~\cite{Neill:2013wsa,Cheung:2018wkq,Bern:2019nnu,Bern:2019crd,Damgaard:2019lfh,Aoude:2020onz,Bern:2021dqo,Bern:2021yeh}, and deducing eikonal-type kernels~\cite{Brandhuber:2021eyq,Kol:2021jjc}.
Progress on scattering amplitudes has also reinvigorated the study of gravitational interactions in a fully (special) relativistic formalism, see for example references~\cite{Damour:2017zjx,Damour:2019lcq,Kalin:2020fhe,Bini:2020uiq,Kalin:2020lmz,Damour:2020tta,Dlapa:2021npj,Bini:2021gat}.

In the next section, we review the observables-based formalism
of refs.~\cite{Kosower:2018adc,Cristofoli:2021vyo}.  We show
how to apply it to the computation of the change of momentum
(equivalent to the scattering angle) in \sect{Impulse}.
In \sect{Waveforms}, we apply it to the waveform of radiation
emitted in the scattering of two particles.  Spinorial
variables are widely used in modern approaches to scattering
amplitudes, and find a natural partner in Newman--Penrose
scalars, which we present in \sect{sec:NP}.
In \sect{sec:continuation}, we review the use of the
$(2,2)$ signature, using it to show the connection
of three-point amplitudes to the linearised field strength or curvature of classical solutions.
In \sect{sec:classicaldc}, we discuss the extension of the amplitudes double copy to a direct relation between classical solutions.
We conclude with some perspectives on future research
in \sect{Conclusions}.

This review is one of a series describing the current status of research in scattering amplitudes and their applications; see ref.~\cite{SAGEX-22-01} for the overview article. Two chapters in this series are closely related to ours: ref.~\cite{SAGEX-22-03}, which reviews the double copy of scattering amplitudes and their applications, and contains a section on classical gravity; and ref.~\cite{SAGEX-22-13}, which is dedicated to the applications of amplitudes in classical gravity, via different approaches to the one we review here.

\section{Observables from Amplitudes}
\label{Observables}

We seek a formalism which will express classical observables
in terms of scattering amplitudes.  We will consider only
observables in scattering processes in this review. There 
are two starting
ingredients in building such a formalism: a parameter
that controls the classical limit, and an expression for
the initial state.

The easiest parameter with which to control the classical
limit is $\hbar$.  We restore it in expressions by
dimensional analysis.  We continue to use relativistic
units, with $c=1$.  We must now distinguish between
units of energy and inverse length ($[M]$ and $[L]^{-1}$ 
respectively).
We normalize the annihilation and creation operators 
for massive scalars so that
\begin{equation}
[\annihilation{p}, \creation{p'}] = 
(2\pi)^3 2E_p \delta^{(3)}(\v{p} - \v{p}')\,,
\label{CreationAnnihilationNormalization}
\end{equation}
with bold symbols denoting spatial three-vectors.
It will be convenient to keep the dimension of single-particle
plane-wave states, $\keta{p}$, to be $[M]^{-1}$ just as
with natural units,
\begin{equation}
\keta{p} \equiv \, \creation{p} \keta{0}\,,
\label{MassiveSingleParticleState}
\end{equation}
with the
vacuum state being dimensionless.  
The state $|p\rangle$
represents a particle of momentum $p$ and positive energy, while $\braa{p}=\braa{0}\annihilation{p}$
is the conjugate state.
We define $n$-particle plane-wave states as simply
the tensor product of normalized single-particle states.

It will be convenient to hide factors of $2\pi$ throughout;
we do that by defining 
an $n$-fold Dirac $\delta$ distribution with normalization
absorbing $2\pi$s,
\begin{equation}
\del^{(n)}(p) \equiv (2\pi)^n \delta^{(n)}(p)\,,
\label{delDefinition}
\end{equation}
where as usual we omit the superscript for $n=1$.
We do the same for the measure,
\begin{equation}
\dd^4 p \equiv \frac{d^4 p}{(2\pi)^4}\,.
\end{equation}
It's convenient as well to define a short-hand notation for
the on-shell phase-space measure,
\begin{equation}
\dPhi(p_i^\vpt) \equiv \dd^4 p_i^\vpt \, \delp(p_i^2-m_i^2)\,.
\label{dfDefinition}
\end{equation}
In this expression,
\begin{equation}
\delp(p^2-m^2) = 2\pi\Theta(p^\timecomponentlabel)\,\delta(p^2-m^2)\,.
\end{equation}
(The energy component of the four-vector is
$p^\timecomponentlabel$.)
We leave the mass implicit in $\dPhi(p)$.

With our normalization of single-particle states,
their inner product is,
\begin{equation}
\langle p' | p \rangle = \Del(p^\vpp-p')\,,
\label{MomentumStateNormalization}
\end{equation}
where,
\begin{equation}
\Del(p^\vpp-p') \equiv  
2 E_{p} \del^{(3)} (\v{p}^\vpp-\v{p}')\,.
\end{equation}
We should understand the argument on the left-hand side as a 
function of four-vectors.
With this abbreviated
notation, we can also rewrite the normalization of 
creation and annihilation
operations~(\ref{CreationAnnihilationNormalization}) in a natural form,
\begin{equation}
[\annihilation{p\vphantom{p'}}, \creation{p'}] = \Del(p - p')\,.
\label{CompactCreationAnnihilationNormalization}
\end{equation}
We will employ the notation 
$\annihilation{}(k)\equiv \annihilation{k}$ and
$\creation{}(k)\equiv \creation{k}$ to allow for additional indices.

The scattering matrix $S$ and the transition matrix $T$ are 
both dimensionless.
Scattering amplitudes are matrix elements of the latter 
between plane-wave states,
\begin{equation}
\langle p'_1 \cdots p'_m | T | p_1 \cdots p_n \rangle = 
\Ampl(p_1 \cdots p_n \rightarrow p'_1 \cdots p'_m) 
\del^{(4)}(p_1 + \cdots p_n - p'_1 - \cdots - p'_m)\,.
\end{equation}
A straightforward consequence of this expression is that $n$-point scattering amplitudes
continue to have mass dimension $[M]^{4-n}$ in our units where $\hbar \neq 1$.

Factors of $\hbar$ appear in two places: in the couplings, and
in the distinction between momenta and wavenumbers.  We consider
theories with massless force carriers generally, denoting the
coupling by $g$.  
In electrodynamics, the coupling is $e$, while
in gravity it is $\kappa=\sqrt{32\pi G}$.  
Amplitudes are power series in $g/\sqrt{\hbar}$, which
introduces an increasing number of inverse powers of $\hbar$ 
with multiplicity and loop order.
An $n$-point $L$-loop amplitude scales as
\begin{equation}
    \hbar^{1-n/2-L}\,.
\end{equation}
The singular powers of $\hbar$ may appear disturbing at
first glance; but we must remember that amplitudes are not
themselves physically observable, and we cannot simply take
the classical limit of amplitudes alone.

In traditional
approaches to quantum field theory, momenta serve as primary variables.
(In more modern approaches, it is rather spinor variables ---
in a sense `square roots' of momenta --- that serve as fundamental
variables.)  We typically make no fundamental
distinction between massive or massless
momenta, and no distinction between momenta and wavenumbers. 
For our purposes, we must
now make this distinction.  The natural variables turn out
to be the momenta of massive particles, but the \textit{wavenumbers\/}
of massless particles (or their spinorial `square roots').

We introduce a notation for the
wavenumber $\wn p$ associated to the momentum $p$,
\begin{equation}
\wn p \equiv p / \hbar\,.
\label{WavenumberNotation}
\end{equation}
The factors of $\hbar$ that arise from expressing observables
in terms of integration over wavenumbers compensate the 
singular factors mentioned above, and ultimately lead to 
finite values for physical observables in the classical limit.

In the usual application of scattering amplitudes to 
collider physics, we take the initial and final states to be
plane-wave states for both massive and massless particles.
We cannot do this if we want to match to a classical limit of
localized point particles.  Instead we must use a
relativistic wavefunction $\phi(p)$ to build a quantum
state $\keta{\psi_1}_\textrm{in}$ corresponding to one localized particle,
\begin{equation}
\begin{aligned}
\keta{\psi_1}_\textrm{in} &= \int \! \dd^4 p\, \delp(p^2 - m^2) 
\, \phi(p) \keta{p}_\textrm{in}
\\ &= 
\int \! \dPhi(p)\;\phi(p) \keta{p}_\textrm{in}\,.
\end{aligned}
\label{OneParticleInitialState}
\end{equation}
It will be easiest to make this state dimensionless, for which
we need $\phi(p)$ to have dimensions of inverse energy.  This
also allows us to normalize the state to unity, which we
can accomplish by normalizing the wavefunctions,
\begin{equation}
\int \! \dPhi(p)\; |\phi(p)|^2 = 1\,.
\label{WavefunctionNormalization}
\end{equation}
The simplest wavefunctions will localize the momentum $p$
in a neighborhood of the classical four-velocity $u$, with
the localization becoming tighter as $\hbar\rightarrow 0$.

Turning to scattering situations,
we build an incoming two-particle state out of localized states
for each particle, with the impact parameter $b$
introduced through a phase factor,
\begin{equation}
\begin{aligned}
| \psi \rangle_\textrm{in} &= \int \! \dd^4 p_1 \dd^4 p_2 
\, \delp(p_1^2 - m_1^2) 
\delp(p_2^2 - m_2^2) \, \phi_1(p_1) \phi_2(p_2) 
\, e^{i b \cdot p_1/\hbar} \keta{p_1\, p_2}_\textrm{in}
\\ &= 
\int \! \dPhi(p_1,p_2)\;
  \phi_1(p_1) \phi_2(p_2) \, e^{i b \cdot p_1/\hbar} 
  \keta{p_1\, p_2}_\textrm{in}
\\ &\equiv \int \dPhi(p_1, p_2) \, \phi_b(p_1, p_2) 
\keta{p_1, p_2}_\textrm{in} 
  \,.
\end{aligned}
\label{InitialState}
\end{equation}
We have abbreviated $\dPhi(p_1,p_2)\equiv \dPhi(p_1)\dPhi(p_2)$.
The impact parameter $b$ is a four-vector, necessarily space-like.
We take it to be transverse, meaning that $p_1\cdot b = 0 = p_2\cdot b$.
With the normalization condition~(\ref{WavefunctionNormalization})
imposed on both $\phi_1(p)$ and $\phi_2(p)$, the initial
state will also be normalized to unity,
${}_\textrm{in}\langle\psi|\psi\rangle_\textrm{in} =1$.

There are three scales that we must consider in the scattering
of massive particles: their Compton wavelengths, 
$\ell_c^{(i)} = \hbar/m_i$; the packet spread in the wavefunctions
$\phi_i(p)$, which we denote $\ell_w^{(i)}$; 
and the scattering length
determined by the interactions, $\ell_s$.
In order to keep quantum effects small, we must insist that
$\ell_c^{(i)}\ll \ell_w^{(i)}$; in order to treat the scattering as
that of point-like particles, we must also insist that
$\ell_w\ll \ell_s$.  The authors of
ref.~\cite{Kosower:2018adc} discuss the combined `Goldilocks' conditions 
in greater detail.  These conditions imply that we should
integrate over wavenumbers for massless particles --- for
both virtual and real-emission lines --- in order to make
the factors of $\hbar$ explicit before integration,
with the integrations
having weight across unbounded domains.  Integrating over momenta
would instead result in the factors of $\hbar$ emerging only
from performing the integrals, and the integrations essentially
concentrated around zero values for the corresponding momenta.

\def\Observable{\langle O\rangle}
We consider observables which can be expressed as the
integrated change
in a quantity over the course of the scattering.  Such an observable
$\Observable$ is determined in the quantum theory by the
difference between the measurement of the corresponding
operator, $\mathbb{O}$, in the final and initial states,
\begin{equation}
    \Observable = 
    {}_\textrm{out}\braa{\psi}\mathbb{O}\keta{\psi}_\textrm{out}
    -{}_\textrm{in}\braa{\psi}\mathbb{O}\keta{\psi}_\textrm{in}\,.
    \label{GenericObservableI}
\end{equation}
\Eqn{InitialState} gives a concrete expression for the
initial state, once we specify the wavefunctions $\phi_i(p)$;
what about the final state?  

We can rely on the $S$ matrix, viewed as the evolution operator
from the far past to the far future, to obtain an expression
for it,
\begin{equation}
\begin{aligned}
 \keta{\psi}_\textrm{out} &= U(\infty,-\infty)\keta{\psi}_\textrm{in}
 \\ &= S\keta{\psi}_\textrm{in}\,.
 \end{aligned}
\end{equation}
We can then express the integrated observable $\Observable$ as follows,
\begin{equation}
    \Observable = 
    {}_\textrm{in}\braa{\psi}S^\dagger\operator{O} S
       -\operator{O}\keta{\psi}_\textrm{in}\,.
\label{NetObservable}
\end{equation}
We will leave the `in' subscript implicit going forward.
We denote the classical limit of $\Observable$ interchangeably
as $\Observable_{\class}$ and just as plain $O$.

We can understand the link between amplitudes and the final-state value of
the observable by inserting a complete set of states,
\begin{equation}
    {}_\textrm{out}\braa{\psi}\operator{O}\keta{\psi}_\textrm{out}
    = \hspace*{4mm}\sumint{X} \hspace*{3mm}
\;{\cal O}(X)\, \bigl| \langle  X |S| \psi\rangle\bigr|^2\,,
\end{equation}
where the $\lsumint{X}$ symbol indicates a summation over all
states $X$ with suitable quantum numbers as well as an integration
over each state's available phase space, and ${\cal O}(X)$ is the
value of the observable in that state.  We have assumed here that
$\operator{O}$ is diagonalizable, and that the states $X$ are
its eigenstates; but the generalization beyond this assumption is
straightforward.

We can simplify the starting expression~(\ref{NetObservable})
by writing 
the scattering matrix in terms of the transition matrix $T$ via
$S = 1 + i T$, in order to make contact with the usual scattering amplitudes.
The no-scattering (unity) part of the $S$ matrix cancels,
\begin{equation}
    \Observable = 
    \braa{\psi}(-i T^\dagger)\operator{O}
    +i \operator{O} T + T^\dagger \operator{O} T\keta{\psi}\,.
\end{equation}
Using the unitarity of the $S$ matrix, $S^\dagger S=1$,
we see that $-i T^\dagger+i T+T^\dagger T = 0$, so we can rewrite
our observable,
\begin{equation}
    \Observable = 
    i\,\braa{\psi}[\operator{O},T]\keta{\psi}
    +\braa{\psi} T^\dagger [\operator{O}, T]\keta{\psi}\,.
\label{FinalObservable}
\end{equation}
This expression holds to all orders in perturbation theory.
The first term is linear in a scattering amplitude, while
the latter will contain a product of scattering amplitudes
(integrated over phase space), or equivalently, the discontinuity
of a scattering amplitude.  It will start at one higher order
in perturbation theory.  This suggests the labels of ``virtual''
and ``cut'' for the two terms, though the second term includes
contributions with virtual corrections as well.

\section{Impulse}
\label{Impulse}
\subsection{Quantum Observable}

A simple but important observable is the net change in the momentum
of one of the scattered particles in the 
initial state~(\ref{InitialState}).
To define this observable, we place detectors at asymptotically 
large distances pointing at the collision region.
The detectors measure only the momentum of one of the
particles, say particle 1. We assume that these detectors cover all
possible scattering angles. Let $\operator P_i^\mu$ be the momentum 
operator for quantum field~$i$; the change in particle 1's momentum
is then,
\begin{equation}
\begin{aligned}
\langle \Delta p_1^\mu \rangle &=  
\braa{\psi}S^\dagger \, \operator P^\mu_1 \, S \keta{\psi} 
- \braa{\psi}\, \operator P^\mu_1 \,\keta{\psi}
\\ &= 
    i\,\braa{\psi}[\operator P^\mu_1,T]\keta{\psi}
    +\braa{\psi} T^\dagger [\operator P^\mu_1, T]\keta{\psi}\,.
\end{aligned}
\label{ImpulseMaster}
\end{equation}
This observable is called the impulse on particle 1.
It is an on-shell observable, defined in both the quantum and 
the classical theories.   We label the first term $\ImpA$,
and the second term~$\ImpB$.
We can similarly measure the impulse imparted to particle 2, 
\begin{equation}
\langle \Delta p_2^\mu \rangle =  
    i\,\braa{\psi}[\operator P^\mu_2,T]\keta{\psi}
    +\braa{\psi} T^\dagger [\operator P^\mu_2, T]\keta{\psi}\,.
\end{equation}

\def\finalk{r}
\def\initialk{p}
\def\initialkc{p'}

Let us take another step, and re-express \eqn{ImpulseMaster} in
terms of scattering amplitudes.  Substituting in the explicit
form of the initial state~(\ref{InitialState}) into $\ImpA$, we find
\begin{equation}
    \begin{aligned}
\ImpA &= 
\int \! \dPhi(\initialk_1,\initialk_2)
                \dPhi(\initialkc_1,\initialkc_2)\;
   e^{i b \cdot (\initialk_1 - \initialkc_1)/\hbar} \, 
      \phi_1(\initialk_1) \phi_1^*(\initialkc_1) 
      \phi_2(\initialk_2) \phi_2^*(\initialkc_2) 
\\ &\hphantom{\dPhi(\initialk_1)\dPhi(\initialk_2)
                \dPhi(\initialkc_1)\dPhi(\initialkc_2)\;
   e^{i b \cdot (\initialk_1 - \initialkc_1)/\hbar} \,}\hspace*{-5mm}
\times i (\initialkc_1\!{}^\mu - \initialk_1^\mu) \, 
     \langle \initialkc_1 \initialkc_2 | \,T\, |
           \initialk_1 \initialk_2 \rangle\,.
\end{aligned}
\label{eq:defl2}
\end{equation}
In this expression, we label the momenta in one copy of 
the initial state by $\initialk_{1,2}$, and those in the conjugate by
$\initialkc_{1,2}$.  We introduce the momentum differences 
$q_i = \initialkc_i-\initialk_i$,
and then change variables in the integration from the $p_i'$ to the $q_i$. 
In these variables, the matrix element is
\begin{equation}
\begin{aligned}
\langle \initialkc_1 \initialkc_2 | \,T\, | \initialk_1 \initialk_2 \rangle &= 
\Ampl(\initialk_1 \initialk_2 \rightarrow \initialkc_1\,, \initialkc_2)
 \del^{(4)}(\initialkc_1+\initialkc_2-\initialk_1-\initialk_2) 
\\&=
\Ampl(\initialk_1 \initialk_2 \rightarrow \initialk_1 + q_1\,, \initialk_2 + q_2)
 \del^{(4)}(q_1 + q_2)\,,
\end{aligned}
\end{equation}
so that the first term in the impulse is
\begin{equation}
\begin{aligned}
\ImpA &= \int \! \dPhi(\initialk_1,\initialk_2)
                 \dPhi(q_1+\initialk_1,q_2+\initialk_2)\;
\\&\qquad\times 
\phi_1(\initialk_1) \phi_1^*(\initialk_1 + q_1)
\phi_2(\initialk_2) \phi_2^*(\initialk_2+q_2) 
\, \del^{(4)}(q_1 + q_2)
\\&\qquad\times 
  \, e^{-i b \cdot q_1/\hbar} 
\,i q_1^\mu  \, \Ampl(\initialk_1 \initialk_2 \rightarrow 
                    \initialk_1 + q_1, \initialk_2 + q_2)\,
\,.
\end{aligned}
\label{ImpulseGeneralTerm1a}
\end{equation}
As a reminder, the shorthand notation for the phase-space measure is,
\begin{equation}
\dPhi(q_1+p_1) = \dd^4 q_1\; \del\bigl((p_1 + q_1)^2 - m_1^2\bigr)
   \Theta(p_1^0 + q_1^0)\,.
\end{equation} 

We perform the integral over $q_2$ using the four-fold delta function. 
Relabelling $q_1 \rightarrow q$, we obtain,
\begin{equation}
\begin{aligned}
\ImpA= \int \! \dPhi(\initialk_1,\initialk_2) &\dd^4 q  \; 
\del(2\initialk_1 \cdot q + q^2) \del(2 \initialk_2 \cdot q - q^2) 
   \Theta(\initialk_1^0+q^0)\Theta(\initialk_2^0-q^0)\\
&\times  e^{-i b \cdot q/\hbar} \phi_1(\initialk_1) \phi_1^*(\initialk_1 + q)
        \phi_2(\initialk_2) \phi_2^*(\initialk_2-q)
\\& \times  
\,  i q^\mu  \, \Ampl(\initialk_1 \initialk_2 \rightarrow 
                      \initialk_1 + q, \initialk_2 - q)\,.
\label{ImpulseGeneralTerm1}
\end{aligned}
\end{equation}
This expression is valid to all orders in perturbation theory.
It is linear in the amplitude.
The incoming and outgoing momenta in this amplitude
do \textit{not\/} correspond to the 
initial- and final-state momenta of the
scattering process.  Rather, both correspond to the 
initial-state momenta,
as they appear in the wavefunction and in its conjugate.  
From the amplitude's point of view, 
the momentum $q$ appears to be a momentum transfer, but for
the physical scattering process it represents a difference between the
momentum in the wavefunction and the one in the conjugate.  We will
call it a `momentum mismatch'.  We may think of $\ImpA$
as the interference of
a standard amplitude with an interactionless forward scattering,
\begin{equation}
\begin{aligned}
\ImpA &= \int \! \dPhi(\initialk_1,\initialk_2)
                \dPhi(\initialkc_1,\initialkc_2)\;
   e^{i b \cdot (\initialk_1 - \initialkc_1)/\hbar} \, 
      \phi_1(\initialk_1) \phi_1^*(\initialkc_1) 
      \phi_2(\initialk_2) \phi_2^*(\initialkc_2) 
\\ &\hphantom{\dPhi(\initialk_1)\dPhi(\initialk_2)
                \dPhi(\initialkc_1)\dPhi(\initialkc_2)\;
   e^{i b \cdot (\initialk_1 - \initialkc_1)/\hbar} \,}\hspace*{-40mm}
\times i \int \dPhi(\finalk_1)\dPhi(\finalk_2)\;(\finalk_1^\mu-\initialk_1^\mu) \, 
   \langle \initialkc_1 \initialkc_2 | \finalk_1 \finalk_2 \rangle
         \langle \finalk_1 \finalk_2 | \,T\, |\initialk_1 \initialk_2 \rangle
\,.
\end{aligned}
\end{equation}

\begin{figure}
    \centering
\begin{equation}
    \begin{aligned}
\int_{\textrm{onshell}}
e^{-i b \cdot q/\hbar}  \,  iq^\mu \times\!\!\!\!
\begin{tikzpicture}[scale=1.0, 
                baseline={([yshift=-\the\dimexpr\fontdimen22\textfont2\relax]
                    current bounding box.center)},
        ] 
\begin{feynman}
	\vertex (b) ;
	\vertex [above left=1 and 0.66 of b] (i1) {$\phi_1(p_1)$};
	\vertex [above right=1 and 0.33 of b] (o1) {$\phi_1^*(p_1+q)$};
	\vertex [below left=1 and 0.66 of b] (i2) {$\phi_2(p_2)$};
	\vertex [below right=1 and 0.33 of b] (o2) {$\phi_2^*(p_2-q)$};
	
	\begin{scope}[decoration={
		markings,
		mark=at position 0.7 with {\arrow{Stealth}}}] 
		\draw[postaction={decorate}] (b) -- (o2);
		\draw[postaction={decorate}] (b) -- (o1);
	\end{scope}
	\begin{scope}[decoration={
		markings,
		mark=at position 0.4 with {\arrow{Stealth}}}] 
		\draw[postaction={decorate}] (i1) -- (b);
		\draw[postaction={decorate}] (i2) -- (b);
	\end{scope}
	
	\filldraw [color=white] (b) circle [radius=10pt];
	\filldraw [fill=allOrderBlue] (b) circle [radius=10pt];
\end{feynman}
\end{tikzpicture} 
\!\!\!\!.
\end{aligned}
\end{equation}
    \caption{A diagrammatic representation of the first term 
    in the impulse~(\ref{ImpulseMaster}), $\ImpA$.}
    \label{Term1Figure}
\end{figure}
\noindent We depict this term diagrammatically in fig.~\ref{Term1Figure}.

Let us next examine the second term $\ImpB$ in the
impulse~(\ref{ImpulseMaster}).  This term contains two transition
matrices, and so we must insert a complete set of states to
connect to scattering amplitudes.  We label these states
by $\finalk_1$, $\finalk_2$ and $X$:
\begin{equation}
\begin{aligned}
\ImpB &= \braa{\psi}\, T^\dagger [ \operator P_1^\mu, T] \,\keta{\psi}
\\&= \hspace*{3mm}\sumint{X} \hspace*{3mm}\int \! \dPhi(\finalk_1) \dPhi(\finalk_2)\; 
\braa{\psi} \, T^\dagger \keta{\finalk_1 \, \finalk_2 \, X} 
\braa{\finalk_1  \, \finalk_2 \, X} [\operator P_1^\mu, T]
\, \keta{\psi}\,.
\end{aligned}
\end{equation}
As before, $X$ contains zero or more messengers.
We again substitute the explicit forms for the
incoming wavefunction~(\ref{InitialState}).  As in $\ImpA$, 
we label the states in $\keta{\psi}$ by $\initialk_{1,2}$, 
those in the conjugate by $\initialkc_{1,2}$.  We identify,
\begin{equation}
\begin{aligned}
\braa{\psi} \, T^\dagger \keta{\finalk_1 \, \finalk_2 \, X} &=
\Ampl^*(\initialkc_1\,, \initialkc_2 \rightarrow \finalk_1 \,, \finalk_2 \,, \finalk_X)
               \del^{(4)}(\initialkc_1+\initialkc_2 - \finalk_1 - \finalk_2 -\finalk_X)
\,,
\\ \braa{\finalk_1  \, \finalk_2 \, X}  T\, \keta{\psi} &=
\Ampl(\initialk_1\,,\initialk_2 \rightarrow 
                      \finalk_1\,, \finalk_2 \,, \finalk_X)
\del^{(4)}(\initialk_1 + \initialk_2 
                                - \finalk_1 - \finalk_2 -\finalk_X) 
   \,.
\end{aligned}    
\end{equation}
In these expressions, $\finalk_X$ denotes the total momentum carried
by the particles in $X$.
We then obtain,
\begin{equation}
\begin{aligned}
\ImpB
&=
\hspace*{3mm}\sumint{X} \hspace*{3mm}
\int \!\prod_{i = 1, 2}  \dPhi(\finalk_i)  \dPhi(\initialk_i) 
\dPhi(\initialkc_i)
    \; \phi_i(\initialk_i) \phi^*_i(\initialkc_i) 
      e^{i b \cdot (\initialk_1 - \initialkc_1) / \hbar}
(\finalk_1^\mu - \initialk_1^\mu) \\
&\hspace*{5mm}\times \del^{(4)}(\initialk_1 + \initialk_2 
                                - \finalk_1 - \finalk_2 -\finalk_X) 
               \del^{(4)}(\initialkc_1+\initialkc_2 - \finalk_1 - \finalk_2 -\finalk_X)\\
&\hspace*{5mm}\times \Ampl(\initialk_1\,,\initialk_2 \rightarrow 
                      \finalk_1\,, \finalk_2 \,, \finalk_X)
   \Ampl^*(\initialkc_1\,, \initialkc_2 \rightarrow \finalk_1 \,, \finalk_2 \,, \finalk_X) 
               \,.
\end{aligned}
\label{eq:forcedef2}
\end{equation}
We again introduce the momentum differences $q_i=p'_i-p_i$, 
integrate over $q_2$, and relabel $q_1\rightarrow q$ to arrive at,
\begin{equation}
\begin{aligned}
\ImpB
&=
\hspace*{3mm}\sumint{X}{}  \hspace*{3mm}
\int \!\prod_{i = 1,2} \dPhi(\finalk_i)  \dPhi(\initialk_i) 
\dd^4 q\;
\del(2\initialk_1 \cdot q + q^2) \del(2 \initialk_2 \cdot q - q^2) 
\\[-3mm]
&\hspace*{25mm}\times    
\Theta(\initialk_1^0+q^0)\Theta(\initialk_2^0-q^0)
\\&\hspace*{25mm}\times    
\phi_1(\initialk_1) \phi_2(\initialk_2)
    \; \phi^*_1(\initialk_1+q) \phi^*_2(\initialk_2-q) 
      e^{-i b \cdot q / \hbar}
(\finalk_1^\mu - \initialk_1^\mu) \\
&\hspace*{25mm}\times \del^{(4)}(\initialk_1 + \initialk_2 
                                - \finalk_1 - \finalk_2 -\finalk_X) 
\\ &\hspace*{25mm}\times \Ampl(\initialk_1\,, \initialk_2 \rightarrow 
                      \finalk_1\,, \finalk_2 \,, \finalk_X)
\Ampl^*(\initialk_1+q\,, \initialk_2-q \rightarrow \finalk_1 \,,\finalk_2 \,, \finalk_X)
               \,.
\end{aligned} 
\label{ImpulseGeneralTerm2b}
\end{equation}
A final change of variables removes the final-state momenta
$\finalk_i$ in favor of momentum transfers $\xfer_i\equiv r_i-p_i$.
These variables are momenta of (virtual) messengers, and thus
scale differently in the classical limit from the massive-particle
momenta.  In terms of them, the expression above is,
\begin{equation}
\begin{aligned}
\ImpB
=
\hspace*{3mm}\sumint{X} \hspace*{3mm}
\int& \!\prod_{i = 1,2}  \dPhi(\initialk_i) \dd^4 \xfer_i
\dd^4 q\;
\del(2p_i\cdot \xfer_i+\xfer_i^2)\Theta(p_i^0+\xfer_i^0)
\\&\hspace*{5mm}\times
\del(2\initialk_1 \cdot q + q^2) \del(2 \initialk_2 \cdot q - q^2) 
   \Theta(\initialk_1^0+q^0)\Theta(\initialk_2^0-q^0)\\
&\hspace*{5mm}\times \phi_1(\initialk_1) \phi_2(\initialk_2)
    \; \phi^*_1(\initialk_1+q) \phi^*_2(\initialk_2-q) 
\\&\hspace*{5mm}\times 
      e^{-i b \cdot q / \hbar}\,\xfer_1^\mu \;
\del^{(4)}(\xfer_1+\xfer_2+\finalk_X) 
\\ &\hspace*{5mm}\times 
 \Ampl(\initialk_1\,, \initialk_2 \rightarrow 
                      \initialk_1+\xfer_1\,, \initialk_2+\xfer_2 \,, \finalk_X)
\\ &\hspace*{5mm}\times 
 \Ampl^*(\initialk_1+q, \initialk_2-q \rightarrow 
   \initialk_1+\xfer_1 \,,\initialk_2+\xfer_2 \,, \finalk_X)
               \,.
\end{aligned} 
\label{eq:impulseGeneralTerm2}
\end{equation}

\begin{figure}
    \centering
\begin{equation}
    \begin{aligned}
\int_{\textrm{onshell}}
\; e^{-i b \cdot q / \hbar}\,\xfer_1^\mu \; \del^{(4)}(\xfer_1+\xfer_2+\finalk_X)  
\begin{tikzpicture}[scale=1.0, 
                baseline={([yshift=-\the\dimexpr\fontdimen22\textfont2\relax]
                    current bounding box.center)},
        ] 
\begin{feynman}
\begin{scope}
	\vertex (ip1) ;
	\vertex [right=3 of ip1] (ip2);
	\node [] (X) at ($ (ip1)!.5!(ip2) $) {};
	\begin{scope}[even odd rule]
	\begin{pgfinterruptboundingbox} % useful to avoid the rectangle in the bounding box
%	\path[invclip] ($  (X) - (4pt, 35pt) $) rectangle ($ (X) + (4pt,35pt) $) ;
	\end{pgfinterruptboundingbox} 

	\vertex [above left=0.66 and 0.5 of ip1] (q1) {$ \phi_1(p_1)$};
	\vertex [above right=0.66 and 0.33 of ip2] (qp1) {$ \phi^*_1(p_1 + q)$};
	\vertex [below left=0.66 and 0.5 of ip1] (q2) {$ \phi_2(p_2)$};
	\vertex [below right=0.66 and 0.33 of ip2] (qp2) {$ \phi^*_2(p_2 - q)$};
	
	\diagram* {
		(ip1) -- [photon]  (ip2)
	};
	\begin{scope}[decoration={
		markings,
		mark=at position 0.4 with {\arrow{Stealth}}}] 
		\draw[postaction={decorate}] (q1) -- (ip1);
		\draw[postaction={decorate}] (q2) -- (ip1);
	\end{scope}
	\begin{scope}[decoration={
		markings,
		mark=at position 0.7 with {\arrow{Stealth}}}] 
		\draw[postaction={decorate}] (ip2) -- (qp1);
		\draw[postaction={decorate}] (ip2) -- (qp2);
	\end{scope}
	\begin{scope}[decoration={
		markings,
		mark=at position 0.34 with {\arrow{Stealth}},
		mark=at position 0.7 with {\arrow{Stealth}}}] 
		\draw[postaction={decorate}] (ip1) to [out=90, in=90,looseness=1.2] node[above left] {{$p_1 + w_1$}} (ip2);
		\draw[postaction={decorate}] (ip1) to [out=270, in=270,looseness=1.2]node[below left] {$p_2 + w_2$} (ip2);
	\end{scope}

	\node [] (Y) at ($(X) + (0,1.4)$) {};
	\node [] (Z) at ($(X) - (0,1.4)$) {};
	\node [] (k) at ($ (X) - (0.65,-0.25) $) {$\finalk_X$};

	\filldraw [color=white] ($ (ip1)$) circle [radius=8pt];
	\filldraw  [fill=allOrderBlue] ($ (ip1) $) circle [radius=8pt];
	
	\filldraw [color=white] ($ (ip2) $) circle [radius=8pt];
	\filldraw  [fill=allOrderBlue] ($ (ip2) $) circle [radius=8pt];
	
\end{scope}
\end{scope}
	  \draw [dashed] (Y) to (Z);
\end{feynman}
\end{tikzpicture} 
\!\!\!\!.
\end{aligned}
\end{equation}
    \caption{A diagrammatic representation of the second term 
    in the impulse~(\ref{ImpulseMaster}), $\ImpB$.}
    \label{Term2Figure}
\end{figure}
We can interpret this term in the impulse as the weighted cut of an
amplitude; we depict it diagrammatically in fig.~\ref{Term2Figure}. 

\subsection{Classical Limit}
We are now ready to take the classical limit.  In order to do
so, we must focus on a particular theory and a particular
order in perturbation theory.  We will study the leading-order
impulse in electrodynamics and in dilaton gravity.
  The two computations are very closely related, and offer
a simple application of the double copy.  

In practice, taking the classical limit requires the following steps.
\begin{itemize}
    \item Replace couplings $g$ by $g/\sqrt{\hbar}$;
    \item Change all messenger momentum variables --- 
    mismatch, virtual, or real-emission --- to wavenumber variables,
    $k = \hbar \kb$;
    \item Approximate $\phi(p+\hbar \qb)$ by $\phi(p)$.  In practice,
    this trivialises all massive-momentum integrals;
    \item Laurent-expand all integrands in $\hbar$;
    \item Replace all massive-particle momenta $p_i$
      by their classical values, $m_i u_i$.
\end{itemize}
It is the change to messenger wavenumber variables, along with the
Goldilocks relation, that justifies the approximation of the
wavefunction $\phi(p+\hbar \qb)$, as it will vary little on
the scale of the Compton wavelength.

We denote the ensemble of these steps via the following notation,
\begin{equation}
\Lexp f(\initialk_1,\initialk_2,\ldots) \Rexp \equiv 
\int \dPhi(\initialk_1,\initialk_2)\;|\phi_1(\initialk_1)|^2\,|\phi_2(\initialk_2)|^2\,
  f(\initialk_1,\initialk_2,\ldots) \,.
\label{AngleBrackets}
\end{equation}
In this expression, $f(\ldots)$ is typically a weighted integral
over an amplitude or product of amplitudes.

The classical limit of the first term in the 
impulse~\eqref{ImpulseGeneralTerm1} 
then takes
the following form to leading order,
\begin{equation}
\begin{aligned}
\ImpAcl = i \Lexp \int \!\dd^4 q  \; 
&\del(2\initialk_1 \cdot q + q^2) \del(2 \initialk_2 \cdot q - q^2) 
   \Theta(\initialk_1^0+q^0)\Theta(\initialk_2^0-q^0)\\
& \times  e^{-i b \cdot q/\hbar} 
\, q^\mu  \, \Ampl^{(0)}(\initialk_1 \initialk_2 \rightarrow 
                      \initialk_1 + q, \initialk_2 - q)\,\Rexp\,.
\end{aligned}
\label{ImpulseGeneralTerm1recast}
\end{equation}
Performing the coupling replacements and the change of
variables detailed above, we obtain,
\begin{equation}
\begin{aligned}
\ImpAclsup{(0)} = 
i \frac{g^2}{4} \Lexp \hbar^2 \int \!\dd^4 \qb  \; 
&\del(\qb\cdot p_1) \del(\qb\cdot p_2) 
\\& \hspace*{-5mm}\times  
e^{-i b \cdot \qb} 
\, \qb^\mu  \, \AmplB^{(0)}(p_1,\,p_2 \rightarrow 
                      p_1 + \hbar\qb, p_2 - \hbar\qb)\,\Rexp\,.
\end{aligned}
\label{ImpulseGeneralTerm1classicalLO}
\end{equation}
At this order, this is the complete result for the
impulse, as $\ImpB$ only enters at ${\cal O}(g^4)$.
In this expression, $\,\AmplB^{(L)}$ denotes
the $L$-loop amplitude with a factor of $g/\sqrt{\hbar}$ 
removed
for every interaction; $\AmplB^{(0)}$ thus denotes the tree-level
amplitude with a factor of $g^2/\hbar$ removed.  
In the electromagnetic case,
this removes a factor of $e/\sqrt{\hbar}$ for each
interaction, while
in the gravitational case, we would remove a
factor of $\kappa/\sqrt{\hbar}$.  In pure electrodynamics or 
gravitational theory, $\,\AmplB$ is independent of couplings.  
It is still $\hbar$-dependent through its dependence
on the momentum mismatch (or more generally, on
momentum transfers).  

We are allowed to drop the $\hbar \qb^2$
terms inside the delta functions 
in \eqn{ImpulseGeneralTerm1classicalLO}
because there are no multiplying
singularities at this order.  At higher orders, this
is no longer true; individual contributions are singular, and
we must expand the delta functions in a power series.  The
singularities in $\hbar$ will cancel in a physical observable
(though there is as yet no all-orders proof of this cancellation). In ref.~\cite{Kosower:2018adc}, this cancellation was checked explicitly at order ${\mathcal O}(G^2)$, and in ref.~\cite{Bern:2021dqo}, the cancellation was checked implicitly to order ${\mathcal O}(G^4)$.

\begin{figure}
    \centering
\begin{tikzpicture}[scale=1.0, baseline={([yshift=-\the\dimexpr\fontdimen22\textfont2\relax] current bounding box.center)}, decoration={markings,mark=at position 0.6 with {\arrow{Stealth}}}]
\begin{feynman}
	\vertex (v1);
	\vertex [below = 1 of v1] (v2);
	\vertex [above left=0.05 and 0.8 of v1] (i1) {$p_1$};
	\vertex [above right=0.05 and 0.8 of v1] (o1) {$p_1+\hbar \qb$};
	\vertex [below left=0.05 and 0.8 of v2] (i2) {$p_2$};
	\vertex [below right=0.05 and 0.8 of v2] (o2) {$p_2-\hbar \qb$};
	\draw [postaction={decorate}] (i1) -- (v1);
	\draw [postaction={decorate}] (v1) -- (o1);
	\draw [postaction={decorate}] (i2) -- (v2);
	\draw [postaction={decorate}] (v2) -- (o2);
	\diagram*{(v1) -- [photon] (v2)};
\end{feynman}	
\end{tikzpicture}
    \caption{The diagram contributing to the $2\rightarrow2$ 
    scattering amplitude.  In electrodynamics, the exchanged line
    is a photon; in dilaton gravity, either a graviton or a dilaton.}
    \label{2to2ScatteringAmplitude}
\end{figure}
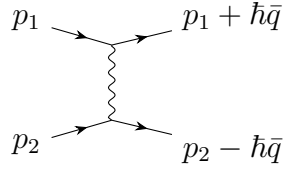

The required $2\rightarrow 2$ scattering amplitude is
given by the lone diagram shown in fig.~\ref{2to2ScatteringAmplitude}.
Its evaluation in electrodynamics gives,
\begin{equation}
\AmplB^{(0)}(p_1 p_2 \rightarrow p_1+\hbar\qb\,, p_2-\hbar\qb) =
 Q_1 Q_2 \frac{4 p_1 \cdot p_2 + \hbar^2 \qb^2}
                                                 {\hbar^2 \qb^2}\,,
\label{ReducedAmplitude1}
\end{equation}
where $Q_{1,2}$ are the charges of the scattering massive 
particles.
The second term is contact-like and becomes irrelevant in the
classical limit.
Substituting this expression into \eqn{ImpulseGeneralTerm1classicalLO}, we obtain
a manifestly $\hbar$-independent expression,
\begin{equation}
\begin{aligned}
\ImpAclsup{(0)} = i e^2 Q_1 Q_2 \Lexp \int \!\dd^4 \qb  \; 
&\del(\qb\cdot p_1) \del(\qb\cdot p_2) 
e^{-i b \cdot \qb} \frac{p_1 \cdot p_2}{\qb^2}
\, \qb^\mu\,\Rexp\,.
\end{aligned}
\label{ImpulseClassicalLOa}
\end{equation}
Applying the remaining actions for the classical limit 
($p_i\rightarrow m_i u_i$), we find
for the impulse,
\begin{equation}
\begin{aligned}
\DeltaPlo = i e^2 Q_1 Q_2 u_1 \cdot u_2\,\int \!\dd^4 \qb  \; 
&\del(\qb\cdot u_1) \del(\qb\cdot u_2) 
e^{-i b \cdot \qb} \frac{1}{\qb^2}\, \qb^\mu\,.
\end{aligned}
\label{ImpulseClassicalLO}
\end{equation}
Evaluating the $\qb$ integral and recognizing
$\gamma = u_1\cdot u_2$ as a Lorentz factor gives us a simple expression,
\begin{equation}
\DeltaPlo  
= -\frac{e^2 Q_1 Q_2}{2\pi} \frac{\gamma}{\sqrt{\gamma^2 - 1}} \frac{b^\mu}{b^2}\,.
\end{equation}

We can use this lowest-order calculation to display a simple
example of the Yang--Mills/gravity double copy in action.  
Following the BCJ form~\cite{Bern:2008qj}
of the double
copy, we obtain an amplitude in a gravitational theory. Squaring the numerator in \eqref{ReducedAmplitude1} leads to, 
\begin{equation}
{\mathcal{\overline {\kern-3pt M}}}^{(0)}(p_1 p_2 \rightarrow p_1+\hbar\qb\,, p_2-\hbar\qb) =
-\frac{1}{16} \frac{\bigl[4 p_1 \cdot p_2\bigr]^2 + {\cal O}(\hbar^2)}
                                                 {\hbar^2 \qb^2}\,.
\label{ReducedAmplitude2}
\end{equation}
This is the amplitude in a theory where the massive particles couple equally to the graviton and to a massless dilaton. We included a factor $-1/16$ which arises in the double copy~\cite{Bern:2010ue,Maybee:2019jus}.
For the pure gravitational theory, in this example, the double copy would require a more involved procedure to project out the dilaton~\cite{Carrasco:2021bmu}. Accordingly, the impulse in the gravitodilatonic scattering
of two massive particles is,
\begin{equation}
\begin{aligned}
\DeltaPlo = -i \frac{\kappa^2}{4} m_1 m_2 (u_1 \cdot u_2)^2\,\int \!\dd^4 \qb  \; 
&\del(\qb\cdot u_1) \del(\qb\cdot u_2) 
e^{-i b \cdot \qb} \frac{1}{\qb^2}\, \qb^\mu\,.
\end{aligned}
\label{ImpulseClassicalLOgravity}
\end{equation}
The integral that appears is identical to the one in
\eqn{ImpulseClassicalLO}; only the factor in front differs.

The impulse is closely related to the classical interaction potential~\cite{Donoghue:1993eb,Donoghue:1994dn,Donoghue:2001qc,Bjerrum-Bohr:2002fji,Bjerrum-Bohr:2002gqz,%
Neill:2013wsa,Bjerrum-Bohr:2013bxa,Cachazo:2017jef,Bjerrum-Bohr:2018xdl,Cristofoli:2019neg,%
Bjerrum-Bohr:2019kec,Emond:2019crr,Brandhuber:2019qpg,Blumlein:2019zku,Cristofoli:2020uzm,Bjerrum-Bohr:2021vuf,Bjerrum-Bohr:2021din} and the eikonal phase~\cite{Torgerson:1966zz,Abarbanel:1969ek,Levy:1969cr,Amati:1987wq,tHooft:1987vrq,Muzinich:1987in,%
Amati:1987uf,Kabat:1992tb,Saotome:2012vy,Melville:2013qca,Akhoury:2013yua,Luna:2016idw,Ciafaloni:2018uwe,%
DiVecchia:2019myk,KoemansCollado:2019ggb,Bern:2020gjj,Parra-Martinez:2020dzs,DiVecchia:2020ymx,DiVecchia:2021ndb,%
DiVecchia:2021bdo,Adamo:2021rfq} 
which provide alternative routes linking classical physics and scattering amplitudes which are discussed 
in more detail in other chapters of this review~\cite{SAGEX-22-03,SAGEX-22-13}. 
The important physical effects of spin can easily be incorporated in the impulse and related
observables using scattering amplitudes~\cite{Maybee:2019jus} building on a deepening understanding
of amplitudes for massive spinning particles~\cite{Arkani-Hamed:2017jhn,Guevara:2017csg,Guevara:2018wpp,Chung:2018kqs,Johansson:2019dnu,%
Chung:2019duq,Siemonsen:2019dsu,Burger:2019wkq,Arkani-Hamed:2019ymq,Chung:2020rrz,Bautista:2021wfy,%
Chiodaroli:2021eug,Bautista:2021inx,Chen:2021qkk,Aoude:2022trd,Bern:2022kto}.

\section{Waveforms and Local Observables}
\label{Waveforms}

It is very satisfying that the impulse, as we saw in the previous 
section, can be computed from scattering amplitudes.
Closely related observables, such as
the radiated momentum, can be computed by very similar 
means~\cite{Kosower:2018adc}.
In this section, we will show that the gravitational waveform itself is determined by scattering amplitudes~\cite{Cristofoli:2021vyo}, at least in the case where two compact objects scatter off one another.
The waveform is an example of a \emph{local} observable, defined as the expectation value of an operator $\Oop(x)$ which is localised at a spacetime point\footnote{This contrasts with global observables, such as the impulse, which do not depend on a specific point.}. 
Thus, local observables have the structure $\bra{\psi} S^\dagger \Oop(x) S \ket{\psi}$.
While there are many possible choices of operator $\Oop(x)$, particularly relevant ones are expectations of the field strength 
tensor $\Fop_{\mu\nu}(x)$ in electrodynamics, 
and of the (linearised) Riemann curvature tensor 
$\Rop_{\mu\nu\rho\sigma}(x)$ in gravity, treated as an effective 
field theory.

\subsection{Tensorial local operators}

We begin with a detailed study of the electrodynamic quantity $\bra{\psi} S^\dagger \Fop_{\mu\nu}(x) S \ket \psi$. 
The analysis for  $\bra{\psi} S^\dagger \Rop_{\mu\nu\rho\sigma}(x) S \ket \psi$ is completely analogous, differing only by the more involved tensor structure.
We will therefore only quote key results in the gravitational case. We return to units with $\hbar = 1$: factors of $\hbar$ can be restored when needed by dimensional analysis.

It is convenient to work in momentum space. The field strength operator can be obtained by differentiating the photon field operator,
with the result
\[
\Fop_{\mu\nu}(x) = -
%\frac{1}{\sqrt{\hbar}} 
2 \Re \sum_{\hel=\pm} \int \dPhi(k) \, i k_{[\mu} \varepsilon_{\nu]}^{(-\hel)}(k) \, e^{-i k \cdot x} \, a_\eta(k) \,.
\label{eq:fieldStrengthDef}
\]
Its expectation can be written directly as
\[
\bra{\psi} S^\dagger \Fop_{\mu\nu}(x)S \ket{\psi} = -
%\frac{1}{\sqrt{\hbar}} 
2 \Re \sum_{\hel=\pm} \int \dPhi(k) \, i k_{[\mu} \varepsilon_{\nu]}^{(-\hel)}(k) \, e^{-i k \cdot x} \, \bra{\psi} S^\dagger a_\eta(k) S \ket{\psi} \, .
\label{eq:Fexpect}
\]
Our aim is to express this quantity in terms of scattering amplitudes. To do so, we expand the $S$ matrix as $S = 1 + i T$, finding
\[
\bra{\psi} S^\dagger a_\hel(k) S \ket{\psi} &= \bra{\psi} (1 - i T^\dagger) a_\hel(k) (1+ i T) \ket \psi \\
%&= i \bra \psi a_\hel(k) T - T^\dagger a_\hel(k) \ket\psi + \bra\psi T^\dagger a_\hel(k) T \ket \psi \\
&= i \bra \psi a_\hel(k) T \ket \psi + \bra\psi T^\dagger a_\hel(k) T \ket \psi \,.
\]
We used $a_\hel(k) \ket \psi = 0$ (there is no incoming radiation in the initial state) to simplify the bra-ket. The field strength expectation becomes,
\[
\bra{\psi} & S^\dagger \Fop_{\mu\nu}(x) S \ket{\psi} = \\
&-
%\frac{1}{\sqrt{\hbar}} 
2 \Re \sum_{\hel=\pm} \int \dPhi(k) \, i k_{[\mu} \varepsilon_{\nu]}^{(-\hel)}(k) \, e^{-i k \cdot x} \, \left[i \bra \psi a_\hel(k) T \ket \psi + \bra\psi T^\dagger a_\hel(k) T \ket \psi \right] \, .
\]
Notice that --- as is true for the impulse discussed in the
previous section ---
the observable inevitably has the structure of two terms: one is linear in $T$ while the second involves two transition matrices.

At this point, it is useful to expand the incoming state~(\ref{InitialState}),
to find
\[
\bra{\psi} S^\dagger \Fop_{\mu\nu}(x) S \ket{\psi} = \hspace*{-2mm}&
\\
-
%\frac{1}{\sqrt{\hbar}} 
2 \Re \sum_{\hel=\pm} \int &\dPhi(k, p_1', p_2', p_1, p_2) \, i k_{[\mu} \varepsilon_{\nu]}^{(-\hel)}(k) \, e^{-i k \cdot x} \, \phi_b^*(p_1', p_2') \phi_b(p_1, p_2) \\
&  \times \left[i \bra{p_1' p_2' k^\hel} T \ket{p_1, p_2} + \bra{p_1' p_2'} T^\dagger a_\hel(k) T \ket {p_1 p_2} \right] \, .
\]
The expectation thus depends on two matrix elements: 
the first, $\bra{p_1' p_2' k^\hel} T \ket{p_1, p_2}$, involves a five-point amplitude while the second, 
$\bra{p_1' p_2'} T^\dagger a_\hel(k) T \ket {p_1 p_2}$, can be written in terms of a product of amplitudes by introducing a sum over states. So indeed
the field strength is determined by scattering amplitudes.

We can expand the amplitudes perturbatively to arrive at an explicit expression. Let us work at leading order. 
We may then ignore $\bra{p_1' p_2'} T^\dagger a_\hel(k) T \ket {p_1 p_2}$: this is a product of two amplitudes, so it is higher order in perturbation theory.
Meanwhile, at lowest order, $\bra{p_1' p_2' k^\hel} T \ket{p_1, p_2}$ is a five-point amplitude of order $g^3$. The leading-order expectation value is,
\[
\bra{\psi} S^\dagger &\Fop_{\mu\nu}(x) S \ket{\psi} = \\
%\frac{1}{\sqrt{\hbar}} 
& 2 \Re \sum_{\hel=\pm}\int \dPhi(k, p_1', p_2', p_1, p_2) \, k_{[\mu} \varepsilon_{\nu]}^{(-\hel)}(k) \, e^{-i k \cdot x} \, \phi_b^*(p_1', p_2') \phi_b(p_1, p_2)\\
& \hspace{100pt} \times  \mathcal{A}(p_1 p_2 \rightarrow p_1' p_2' k^\hel) \del(p_1 + p_2 - p_1' - p_2' -k) \,.
\label{eq:loFieldStrengthQuantum}
\]

Despite this derivation, it may seem contrary to intuition that the classical radiation field is 
related directly to a five-point amplitude.
This amplitude describes the emission of \emph{one} photon, while 
clearly the classical field must contain many photons.
However there is no contradiction, as discussed in detail in
reference~\cite{Cristofoli:2021jas}. 
Classical radiation is described by a coherent state
of the form 
\[
\ket{\alpha} = \mathcal{N}_\alpha
\exp \left[ \sum_\hel \int \dPhi(k) \, \alpha_\eta(k) a_\hel^\dagger(k) \right] \ket{0} \,,
\]
where the function $\alpha_\hel(k)$ parameterises the
radiation, and $\mathcal{N}_\alpha$ is a normalisation
factor.
The expectation of the field-strength operator on such
a coherent state is
\[
\bra{\alpha} \Fop_{\mu\nu}(x) \ket{\alpha}
=
-
2 \Re \sum_{\hel=\pm} \int \dPhi(k) \, i k_{[\mu} \varepsilon_{\nu]}^{(-\hel)}(k) \, e^{-i k \cdot x} \, \alpha_\eta(k) \,.
\]
Comparing to equation~\eqref{eq:loFieldStrengthQuantum}, 
we see that the parameter $\alpha$  of the coherent state is
determined by the five-point amplitude.
It is straightforward to check that the expectation value of the occupation number of the coherent state
is large in the classical limit, as one would expect.

So far, we have not taken advantage of simplifications available 
in the classical limit. 
This is manifest in equation~\eqref{eq:loFieldStrengthQuantum} 
in the dependence on the incoming wavepacket.
Clearly classical quantities cannot depend on the details of the 
wavepacket; we must be able to take advantage of general properties
of wavepackets of particles in the classical regime to remove the 
explicit wavepackets from our expressions. 
As a first step, we again introduce momentum mismatches $q_1$ 
and $q_2$ by setting,
\[
p_1' = p_1 - q_1 \,, \quad
p_2' = p_2 - q_2 \,.
\]
Conservation of momentum then tells us that $k = q_1 + q_2$. As usual~\cite{Kosower:2018adc}, the momentum mismatches are of order $\hbar$. 
In terms of these variables, the on-shell phase space measure in equation~\eqref{eq:loFieldStrengthQuantum} can be written as
\[
\int \dPhi(k, p_1', p_2', p_1, p_2) = \int d\Phi(k, p_1, p_2) \int \dd^4 q_1 \dd^4 q_2 \, \del(2p_1 \cdot q_1 - q_1^2) \del(2 p_2 \cdot q_2 - q_2^2) \,,
\]
up to Heaviside theta functions $\Theta(p_1^0 - q_1^0) \Theta (p_2^0 - q_2^0)$. 
These Heaviside functions are always unity in the classical region, because the scattering must be well below the pair-production threshold: $q_i^0 \ll p_i^0$.

In terms of the momentum mismatches, the wavepackets in equation~\eqref{eq:loFieldStrengthQuantum} become 
\[
\phi_b^*(p_1 - q_1, p_2 -q_2) \phi_b(p_1, p_2) &= 
\phi^*(p_1 - q_1)\phi^*(p_2 -q_2) \phi(p_1)\phi(p_2) 
e^{i q_1 \cdot b /\hbar} \\
&\simeq |\phi(p_1)\phi(p_2)|^2 e^{i q_1 \cdot b /\hbar} \,.
\]
In the last line, we took advantage of the fact that 
the wavefunctions vary little on
the scale of the Compton wavelength.

We can now write the field strength expectation as
\[
\bra{\psi} S^\dagger \Fop_{\mu\nu}(x)S \ket{\psi} = \hspace*{1mm}&
\\
%\frac{1}{\sqrt{\hbar}} 
2 \Re \sum_{\hel=\pm} & 
\int \dPhi(k, p_1, p_2) \,|\phi(p_1)\phi(p_2)|^2 
\int \dd^4 q_1 \dd^4 q_2 \, 
\del(2 p_1 \cdot q_1) \del(2 p_2 \cdot q_2)
\\
& \times k_{[\mu} \varepsilon_{\nu]}^{(-\hel)}(k) \, e^{i(q_1 \cdot b- k \cdot x)} \, \mathcal{A}(p_1 p_2 \rightarrow p_1' p_2' k^\hel) \, \del(q_1 + q_2 -k) \,.
\]
We have approximated $\del(2p_i \cdot q_i - q_i^2) \simeq \del(2p_i \cdot q_i)$; the $q_i^2$ shift is negligible in this leading-order perturbative computation.

The basic role of the wavepackets now is to set the initial conditions: namely, that the relevant momenta $p_i$ in the integral are all equal to initial classical values, up to some spread of order the size of the wavepacket. At leading order 
in the coupling, we are entitled to take this spread to be negligible;
we could simply evaluate the $p_1$ and $p_2$ integrals.
At higher orders, the situation is more complicated. 
Excess inverse powers of $\hbar$ cancel between higher order (loop) 
five-point amplitudes and 
the $\bra{p_1' p_2'} T^\dagger a_\hel(k) T \ket {p_1 p_2}$ term. 
It is only after this cancellation takes place that we can extract 
a classical result. 
We therefore use the notation~\eqref{AngleBrackets}
to tidy away the remaining integrals over $p_1$ and $p_2$, 
as well as the wavefunction. 
In this way, we obtain,
\[
\bra{\psi} S^\dagger \Fop_{\mu\nu}(x) S \ket{\psi} = \hspace*{0mm}&
\\
2 \Re \sum_{\hel=\pm} \Lexp \int &\dPhi(k) \, \dd^4 q_1 \dd^4 q_2 \, \del(2 p_1 \cdot q_1) \del(2 p_2 \cdot q_2) \, \del(q_1 + q_2 -k)
\\[-2mm]
& \times k_{[\mu} \varepsilon_{\nu]}^{(-\hel)}(k) \, e^{i(q_1 \cdot b- k \cdot x)} \, \mathcal{A}(p_1 p_2 \rightarrow p_1' p_2' k^\hel) \Rexp \,.
\label{eq:fsIntegral}
\]

In the gravitational case, assuming that the spacetime is asymptotically Minkowskian, we measure the expectation of the Riemann curvature rather than of the field strength. 
We are only interested in this expectation value at (literally) astronomical distances $r$ from the source of gravitational waves
\footnote{It may be interesting to consider situations where
other gravitating objects are also present. This is
beyond the scope of this review.}
, so we are entitled to assume
that the metric is of the form
\[
g_{\mu\nu}(x) = \eta_{\mu\nu} + \kappa h_{\mu\nu}(x) \,,
\]
where $\kappa h_{\mu\nu}(x)$ is small compared to the Minkowski metric $\eta_{\mu\nu}$. 
%We choose
%\[
%\kappa = \sqrt{32 \pi G}\,,
%\]
%where $G$ is Newton's constant. 
The deviation $h_{\mu\nu}$ has an expansion in distance $r$ beginning at $1/r$; only this term is of relevance. 
For this reason, we are entitled to restrict attention to the linearised Riemann curvature operator, which (in our conventions) is explicitly given by
\[
\Rop_{\mu\nu\rho\sigma}(x) = \frac{\kappa}{2} 
%\frac{1}{\sqrt{\hbar}} 
2 \Re \sum_{\hel = \pm} \int \dPhi(k) \, k_{[\mu} \varepsilon_{\nu]}^{(-\hel)}(k)  \, k_{[\rho} \varepsilon_{\sigma]}^{(-\hel)}(k) \, e^{-i k \cdot x} \, a_\eta(k) \,,
\label{eq:Riemannop}
\]
which can be compared to the field strength operator of equation~\eqref{eq:fieldStrengthDef}. 
The expectation of this operator can be computed in complete parallel with the electromagnetic case, resulting in
\[
\bra{\psi} S^\dagger \Rop_{\mu\nu\rho\sigma}(x) S\ket{\psi} =
%\frac{\kappa}{\sqrt{\hbar}} 
\kappa
\Re \sum_{\hel=\pm}& \Lexp \int \dPhi(k) \, \dd^4 q_1 \dd^4 q_2 \, \del(2 p_1 \cdot q_1) \del(2 p_2 \cdot q_2) \, \del(q_1 + q_2 -k)
\\
& \times i k_{[\mu} \varepsilon_{\nu]}^{(-\hel)}(k) \,  k_{[\sigma} \varepsilon_{\rho]}^{(-\hel)}(k) \, e^{i(q_1 \cdot b- k \cdot x)} \, \mathcal{M}(p_1 p_2 \rightarrow p_1' p_2' k^\hel) \Rexp \,.
\label{eq:curvIntegral}
\]

\subsection{Performing the on-shell integral}

Given the five-point tree amplitude $\mathcal{M}(p_1 p_2 \rightarrow p_1' p_2' k^\hel)$, equation~\eqref{eq:curvIntegral} shows that the classical curvature is obtained by performing integrals over the mismatches $q_1$, $q_2$ and also over the wavevector $k$. 
Similarly, equation~\eqref{eq:fsIntegral} expresses the electromagnetic field strength in terms of very similar integrals.

Of these integrals, the on-shell integral over $k$ can be performed, in part, without knowledge of the amplitude.
There are many ways of performing this integral; one method was reviewed carefully in reference~\cite{Cristofoli:2021vyo}. 
In this review, we choose to perform the integral in a slightly more 
direct manner, at the price of making the boundary conditions less 
transparent than
in the lengthier discussion~\cite{Cristofoli:2021vyo}. 

We focus for clarity on the electromagnetic case; once again, the gravitational case is very similar.
Let us write the expectation of the field strength as,
\[
\bra{\psi} S^\dagger \Fop_{\mu\nu}(x) S \ket{\psi} = 2 \Re i \int \dPhi(k) \tilde J_{\mu\nu}(k) e^{-i k \cdot x} \,,
\]
where $\tilde J_{\mu\nu}(k)$ can be determined by comparing to equation~\eqref{eq:fsIntegral}:
\[
\tilde J_{\mu\nu}(k) = 
-
%\frac{i}{\sqrt{\hbar}} 
i
\sum_{\hel=\pm} \Lexp \int \dd^4 q_1 \dd^4 q_2 &\, \del(2 p_1 \cdot q_1) \del(2 p_2 \cdot q_2) \, \del(q_1 + q_2 -k)
\\& \times 
k_{[\mu} \varepsilon_{\nu]}^{(-\hel)}(k) \, e^{iq_1 \cdot b} \, \mathcal{A}(p_1 p_2 \rightarrow p_1' p_2' k^\hel) \Rexp \,.
\label{eq:Jdef}
\]
We write the wavevector as $k = (\omega, \omega \v n)$ in preparation for performing the integral over the spatial components of the wavevector,
extracting a power of the frequency $\omega$ so that we may integrate over a dimensionless vector $\v n$. 
Then the field strength becomes
\[
\bra{\psi} S^\dagger \Fop_{\mu\nu}(x) S \ket{\psi} &= 2 \Re i \int_0^\infty \dd \omega \int \dd^3 n\, \del(1 - \v n^2) \, e^{-i \omega x^0} e^{i \omega \, \v n \cdot \v x} \, \omega \tilde J_{\mu\nu}(\omega, \omega \v n) \\
&= 2 \Re i \int_0^\infty \dd \omega \int \dd^3 n \int d\lambda \, e^{i \lambda (1 - \v n^2)} e^{-i \omega x^0} e^{i \omega \, \v n \cdot \v x}  \, \omega \tilde J_{\mu\nu}(\omega, \omega \v n)\,.
\]
In the second equality, we wrote the delta function which enforces the on-shell constraint as an integral.

We now proceed to perform the three integrals over $n$ \emph{and} the integral over $\lambda$ by stationary phase, 
justified by the presence of the large distance $|\v x|$ in the phase factor.
In general, the stationary phase approximation is
\[
\int \d^n z \, e^{i f(z)} \, g(z) \simeq (2\pi)^{n/2} \sum_{z_0} e^{i f(z_0)} g(z_0)  \frac{1}{\sqrt{|\det H(z_0)|}} e^{\frac{i\pi}{4} \sign H(z_0)} \,,
\]
where $H(z)$ is the Hessian matrix of second partial derivatives of $f(z)$, $\sign H(z)$ is its signature, and the sum runs over all solutions of the stationary phase condition
\[
f'(z_0) = 0 \,.
\]
In our case, $z=(\v n,\lambda)$, and the stationary phase conditions are 
\[
\v n^2 &= 1 \,,\quad
2 \lambda \, \v n &= \omega \, \v x \,.
\]
These can be solved trivially. There are in fact two solutions:
\[
\v n = \frac{\v x}{|\v x|} \,,\;\; \lambda = \frac{\omega}2 |\v x| \,,
\]
and
\[
\v n = -\frac{\v x}{|\v x|}  \, , \;\;\lambda = -\frac{\omega}2 |\v x| \,.
\]
We will soon see that these two branches correspond to retarded and advanced propagation. For now we retain both.

To complete the evaluation of our four integrals by stationary phase, we need the Hessian matrix, which is
\[
H = \begin{pmatrix}
-2 \lambda & 0& 0 & -2 n_x \\
0 & -2 \lambda & 0 & -2 n_y \\
0 & 0 & -2 \lambda & -2 n_z \\
-2 n_x & -2 n_y & -2 n_z & 0 
\end{pmatrix} \,.
\]
The determinant is $-4 \omega^2 |\v x|^2$, and the four eigenvalues are $-2 \lambda, -2 \lambda, -\lambda + \sqrt{4 + \lambda^2}, -\lambda - \sqrt{4 + \lambda^2}$. Thus when $\lambda ={\omega} |\v x|/2$, the signature of the Hessian is $-2$, while when $\lambda = -{\omega}|\v x|/2$ the signature is $+2$. 

Having gathered this information, we can perform the integral, finding
\[
\bra{\psi} S^\dagger \Fop_{\mu\nu}(x) S \ket{\psi} = 2 \Re \frac{1}{4\pi |\v x|} \int_0^\infty \dd \omega &\left[ e^{-i \omega(x^0 - |\v x|)} \tilde J_{\mu\nu}(\omega, \omega \v x / |\v x|) 
\right. \\
& \qquad\left.
-  e^{-i \omega(x^0 + |\v x|)} \tilde J_{\mu\nu}(\omega, -\omega \v x / |\v x|) \right] \,.
\]
Notice that the two terms depend either on the retarded time $x^0 - |\v x|$ or on the advanced time $x^0 + | \v x |$. 
Since we place our observer at a position $x$ in the future of the event which generates radiation, we may ignore the second (advanced) term.
Indeed, in the more careful discussion in reference~\cite{Cristofoli:2021vyo} it is clear that this second term vanishes for observers in the future of the event.

Thus we arrive at
\[
\bra{\psi} S^\dagger \Fop_{\mu\nu}(x) S \ket{\psi} &= 2 \Re \frac{1}{4\pi |\v x|} \int_0^\infty \dd \omega \, e^{-i \omega(x^0 - |\v x|)} \tilde J_{\mu\nu}(\omega, \omega \v x / |\v x|)  \\
&= \frac{1}{4\pi |\v x|} \int_{-\infty}^\infty \dd \omega \, e^{-i \omega(x^0 - |\v x|)} \tilde J_{\mu\nu}(\omega, \omega \v x / |\v x|) \,,
\label{eq:fsIntegrated}
\]
where in the second expression we define $\tilde J_{\mu\nu}$ for negative values of $\omega$ as
\[
\tilde J_{\mu\nu}(- |\omega| , - |\omega| \v x / |\v x|) \equiv \tilde J_{\mu\nu}^*(+|\omega| , + |\omega| \v x / |\v x|) \,, \quad \textrm{ when } \omega < 0 \,.
\]
This ensures that the reality condition is satisfied.

The factor $1/|\v x|$ in our expression~\eqref{eq:fsIntegrated} for the field strength is precisely as expected for a radiation field. The analogue of the gravitational waveform in this electromagnetic case is the coefficient of this trivial $1/|\v x|$ fall-off.
We therefore define a spectral waveform $f_{\mu\nu}$ as,
\[
\bra{\psi} S^\dagger \Fop_{\mu\nu}(x) S \ket{\psi} = \int \dd \omega \, e^{-i\omega(x^0 - |\v x|)} \frac{f_{\mu\nu}(\omega, \v n)}{|\v x|} \,,
\label{eq:fsWaveformDef}
\]
where $\v n = \v x / |\v x|$. The spectral waveform depends on the frequency and the direction $\v n$ from the source to the observer. Notice that we have chosen to Fourier transform with respect to the retarded time $t = x^0 - |\v x|$ for convenience. Referring back to equation~\eqref{eq:Jdef} for the definition of $\tilde J_{\mu\nu}$, we may write the spectral waveform as,
\[
f_{\mu\nu}(\omega, \v n) = \frac{-i}{4\pi}  \sum_{\hel=\pm} \Lexp \int \dd^4 q_1 & \dd^4 q_2\, \del(2 p_1 \cdot q_1) \del(2 p_2 \cdot q_2) \, \del(q_1 + q_2 -k)
\\& \times 
\left. k_{[\mu} \varepsilon_{\nu]}^{(-\hel)}(k) \, e^{iq_1 \cdot b} \, \mathcal{A}(p_1 p_2 \rightarrow p_1' p_2' k^\hel) \Rexp  \right|_{k = \omega (1, \v{n})}\,.
\label{eq:fsSpectral}
\]

In the gravitational case, we may again write,
\[
\bra{\psi} S^\dagger \Rop_{\mu\nu\rho\sigma}(x) S\ket{\psi} &= 
2 \Re i \int \dPhi(k) \, \tilde J_{\mu\nu\rho\sigma}(k) e^{-i k \cdot x} \\
&\simeq 2 \Re \frac{1}{4\pi |\v x|} \int \dd \omega \, e^{-i \omega (x^0 - |\v x|)} \tilde J_{\mu\nu\rho\sigma}(\omega, \omega \v n) \,.
\]
In the second step, we used the stationary phase approximation.
%
%where
%\[
%\tilde J_{\mu\nu\rho\sigma}(k) = 
%-i \frac{\kappa}{2} \frac{1}{\sqrt{\hbar}} \sum_{\hel=\pm} \Lexp& \int \dd^4 q_1 \dd^4 q_2 \, \del(2 p_1 \cdot q_1) \del(2 p_2 \cdot q_2) \, \del(q_1 + q_2 -k)
%\\
%& \times  \kb_{[\mu} \varepsilon_{\nu]}^{(-\hel)}(k) \,  \kb_{[\sigma} \varepsilon_{\rho]}^{(-\hel)}(k) \, e^{iq_1 \cdot b/\hbar} \, \mathcal{M}(p_1 p_2 \rightarrow p_1' p_2' k^\hel) \Rexp \,.
%\]
Defining a spectral gravitational waveform via
\[
\bra{\psi} S^\dagger \Rop_{\mu\nu\rho\sigma}(x) S\ket{\psi} 
= \int \dd \omega \,e^{-i\omega(x^0 - |\v x|)} \frac{f_{\mu\nu\rho\sigma}(\omega, \v n)}{|\v x|} \,,
\label{eq:curvatureWaveformDef}
\]
we finally arrive at an expression for the spectral waveform:
\[
f_{\mu\nu\rho\sigma}(\omega, \v n) = 
\frac{\kappa}{8 \pi} & %\frac{1}{\sqrt{\hbar}} 
\sum_{\hel=\pm} \Lexp \int \dd^4 q_1 \dd^4 q_2 \, \del(2 p_1 \cdot q_1) \del(2 p_2 \cdot q_2) \, \del(q_1 + q_2 -k)
\\
& \times \left. k_{[\mu} \varepsilon_{\nu]}^{(-\hel)}(k) \,  k_{[\sigma} \varepsilon_{\rho]}^{(-\hel)}(k) \, e^{iq_1 \cdot b} \, \mathcal{M}(p_1 p_2 \rightarrow p_1' p_2' k^\hel) \Rexp \right|_{k = \omega (1, \v{n})} \,.
\label{eq:spectralTensorWaveform}
\]

Thus we see that the gravitational waveform is indeed intimately related to five-point amplitudes.
It is interesting to consider the behaviour of the waveform in the soft limit, where the frequency $\omega$ of the outgoing radiation becomes very small. In this limit, the five-point amplitude in
equation~\eqref{eq:spectralTensorWaveform} 
is equal to a four-point amplitude $\mathcal{M}(p_1 p_2 \rightarrow p_1' p_2')$ times the Weinberg
soft factor. 
It is then possible to show that the waveform in
this long-frequency limit
is determined by the impulse~\cite{Strominger:2014pwa,Bautista:2019tdr}.
This phenomenon is an example of the classical
memory effect~\cite{memory1,Braginsky:1985vlg,memory3}. 

As five-point amplitudes are intimately related to classical radiation, they appear in expressions for other classical observables related to radiation.
An important case studied in~\cite{Kosower:2018adc} is the total momentum
radiated during a scattering event. 
Physically this is connected to the total energy flux, an important physical
observable in gravitational wave physics. 
It has been computed using the KMOC formalism at order $G^3$~\cite{Herrmann:2021lqe,Herrmann:2021tct} in gravity,
and at order $\alpha^3$ in electrodynamics~\cite{Bern:2021xze}.
Another interesting observable which connects to five-point amplitudes is the radiated angular momentum~\cite{Manohar:2022dea}. 
The relevance of five-point amplitudes with massive legs to gravitational radiation motivates determining these amplitudes at loop order~\cite{Carrasco:2020ywq,Carrasco:2021bmu}.

\section{Newman--Penrose Scalars}
\label{sec:NP}
% DOC?
% NP scalars / (2,2) signature (as a contour choice) spinor indices / \Phi_2's and \Psi_4's. Include all-order expressions
% Radiative things -> connect to previous section
% What about then the \Phi_1 or Psi_2 / Coulombic: motivates analytic continuation (linearised)

In the previous section, we saw that waveforms (gravitational and the electromagnetic analogue) may be written in terms of scattering amplitudes.
Of course, the actual waveforms measured in observatories are not tensors: they are scalar quantities. 
The relevant scalar is essentially a component of the tensor quantity~\eqref{eq:spectralTensorWaveform}. 
In this section, we will extract the relevant component. 

In a gravitational wave observatory, measurements are made of the strain $h$. In transverse-traceless gauge, the second time derivative of the strain,
$\ddot h$ is given by the Riemann tensor (see for example the textbook~\cite{Maggiore:2007ulw})
\[
\ddot h \sim e^{ij} R_{i0j0} \,,
\]
where $e^{ij}$ is an appropriate polarisation object.
Thus knowledge of the curvature is knowledge of the waveform. 

The Newman-Penrose (NP) formalism~\cite{Newman:1961qr} is convenient for capturing the relevant component of the waveform, 
and (as we will see) is also closely related to the successful helicity-based approach to amplitudes.
One way to introduce the NP formalism is to choose a convenient null tetrad --- that is, a basis of spacetime vectors $L$, $N$, $M$ 
and $\bar M$ satisfying\footnote{These vectors were denoted $l$, $n$, $m$ and $\bar m$ in the original paper~\cite{Newman:1961qr}. 
We use upper case symbols to distinguish the vectors from masses and loop momenta in quantum field theory.}
\[
L \cdot N = 1, \quad M \cdot \bar M = -1 \,,
\]
with all other dot products vanishing. We may choose $L$ and $N$ to be real vectors, while $M$ and $\bar M$ are complex conjugates of one another.

Referring to either the electromagnetic field strength or gravitational curvature computed in the previous section, it is natural to define
\[
L = (1 , \v x / |\v x|) \,,
\]
so that $L$ is the null vector from the source event to the observer.
Then the wavevector $k$ in equations~\eqref{eq:fsSpectral} and~\eqref{eq:spectralTensorWaveform} is $k = \omega L$. 
We can complete the basis by choosing $M$ and $\bar M$ to be polarisation vectors $\varepsilon^{(\hel)}(k)$ associated with the wavevector $k$; 
these do not depend on the frequency $\omega$, so we may regard $M$ and $\bar M$ to be spacetime vectors associated with $L$. This is
helpful since it is useful to commute the NP basis vectors through the frequency integral in equations~\eqref{eq:fsWaveformDef} and~\eqref{eq:curvatureWaveformDef}. 
We complete the definition of the NP tetrad by choosing $N$ to be the gauge vector associated with the polarisations $\varepsilon^{(\hel)}(k)$, scaled
so that $L \cdot N = 1$.
That is,
\[
L = k / \omega\,, \quad M = \varepsilon^{(+)}(k)\,, \quad \bar M = \varepsilon^{(-)}(k) \,, \quad N = \frac12 (1, -\v x / |\v x|) \,,
\]
where we made a particular gauge choice to define $N$:
\[
N \cdot \varepsilon^{(\pm)}(k) = 0 \,.
\]
It follows that the curvature component $R_{i0j0}$ relevant for the gravitational strain can be obtained from knowledge of 
\[
\Psi_4(x) \equiv - R_{\mu\nu\rho\sigma}(x) N^\mu \bar M^\nu N^\rho \bar M^\sigma  
\]
and its complex conjugate. The object $\Psi_4(x)$ is one of the Newman-Penrose scalars associated with the curvature.
In scattering phenomena on asymptotically Minkowski spacetimes,
all other components of the curvature are suppressed relative
to $\Psi_4(x)$ (see for example, the useful textbook~\cite{Stewart:1990uf}) and so we do not consider them.

Indeed, this is consistent with our understanding of the curvature 
at large distances from the source. 
Let us define the spectral form of $\Psi_4^{1}(x)$ as,
\[
\Psi_4^{1}(x) = \int \dd\omega \,e^{-i \omega (x^0 - |\v x|)} \, \tilde \Psi_4^1(\omega, \v n) \,.
\]
Using equations~\eqref{eq:curvatureWaveformDef} and~\eqref{eq:spectralTensorWaveform} we find
\[
\tilde \Psi_4^{1}(\omega, \v n) = 
\frac{-\kappa}{8 \pi} \omega^2 
%\frac{1}{\sqrt{\hbar}}
\sum_{\hel=\pm} \Lexp \int \dd^4 q_1 \dd^4 q_2 \,& \del(2 p_1 \cdot q_1) \del(2 p_2 \cdot q_2) \, \del(q_1 + q_2 -k)
\\
& \times \left.  \, e^{iq_1 \cdot b} \, \mathcal{M}(p_1 p_2 \rightarrow p_1' p_2' k^+) \Rexp \right|_{k = \omega (1, \v{n})}  \,.
\]
The frequency-space NP scalar is an integral of the helicity amplitude.

The situation in electrodynamics is very similar. The electrodynamic NP scalar most comparable to $\Psi_4(x)$ is $\Phi_2(x) = F_{\mu\nu}(x) \bar M^\mu N^\nu$. Again peeling tells us that this component falls off as $1/\textrm{distance}$ while the other electrodynamic NP scalars fall off more rapidly. 
The leading term of $\Phi_2(x)$ in this distance expansion, $\Phi_2^1(x)$, is proportional to a five-point helicity amplitude, this time with a final-state photon 
replacing the final-state graviton~\cite{Cristofoli:2021vyo}.

\section{Curvature in (2,2) signature}
\label{sec:continuation}

In section~\ref{sec:NP}, we saw that the radiation field (that is, the part of the field strength or curvature which falls off as $1/\textrm{distance}$) due to
a scattering event can be recovered from five-point amplitudes. 
An immediate question is whether it is possible to recover static Coulomb/Schwarzschild-type fields from amplitudes.
It is clear that the five-point amplitudes discussed in the previous sections have no role for static fields: these fields only involve one (incoming and outgoing) massive particle!
Instead, three-point amplitudes involving one massive line and one 
messenger (photon or graviton) are the natural starting point.
Because the on-shell conditions for these amplitudes lack
non-trivial support in Minkowski space, 
we instead turn to a spacetime with metric signature $(+, +, -, -)$: 
such spacetimes are variously described as having 
$(2,2)$~\cite{Arkani-Hamed:2009hub}, split~\cite{Mason:2005qu}, or 
Kleinian~\cite{Crawley:2021auj} signature. An alternative viewpoint, which we will not employ here, is to consider complex kinematics in Minkowski spacetime.

One advantage of working in real $(2,2)$ signature (rather than complex kinematics) is that we may work with real spinors: 
as we will see, these provide an alternative route to manifestly gauge-invariant expressions for curvature.
Since we are working in linearised theory, we can use flat-space methods to introduce spinors, even in the gravitational context. 
The basic set up is similar to the standard spinor-helicity formalism in four-dimensional scattering amplitudes.
We introduce the $\sigma$ matrices 
\[
\label{eq:sigmachiral}
\sigma^\mu = (1, i\sigma_y, \sigma_z, \sigma_x)\,, \quad \tilde \sigma = (1, -i \sigma_y, -\sigma_z, -\sigma_x) \,,
\]
where $\sigma_{x,y,z}$ are the Pauli matrices. Note that each $\sigma^\mu$ matrix is real. The generators of the Lorentz algebra in the chiral spinor representation are proportional to the six matrices 
\[
\label{eq:sigmaantichiral}
\sigma^{\mu\nu} = \frac{1}{4} (\sigma^\mu \tilde \sigma^\nu - \sigma^\nu \tilde \sigma^\mu) \,.
\]
Similarly, in the antichiral representation we have
\[
\tilde \sigma^{\mu\nu} = \frac{1}{4} (\tilde \sigma^\mu \sigma^\nu - \tilde \sigma^\nu \sigma^\mu) \,.
\]
The basic reason for the utility of spinors in our discussion is the following. 
We have seen above that the electromagnetic field strength, and the Riemann curvature, involve the
quantity
\[
k_{[\mu} \varepsilon_{\nu]}^{(-)}(k) = -\frac{1}{\sqrt{2}} \bra{k} \sigma_{\mu\nu} \ket{k} \,,
\]
where the chiral spinor $\ket{k}$, and its antichiral compatriot $|k]$, are defined by
\[
k \cdot \sigma = |k] \bra{k} \,.
\]
As a result, we can write manifestly gauge-invariant expressions for the field strength and curvature using spinors.

We define the Weyl spinor as a spinorial form of the curvature\footnote{We assume that the observation point $x$ is in vacuum, so that the Riemann and Weyl curvature tensors are equal.}
\[
\Psi_{\alpha\beta\gamma\delta}(x) = \sigma^{\mu\nu}_{\alpha\beta} \sigma^{\rho\sigma}_{\gamma\delta} R_{\mu\nu\rho\sigma}(x) \,.
\]
This spinor is associated to the self-dual part of the curvature, because it is obtained from the chiral representation \eqref{eq:sigmachiral}. The conjugate Weyl spinor, associated to the anti-self-dual part of the curvature, is obtained by replacing $\sigma^{\lambda\eta}\rightarrow\tilde\sigma^{\lambda\eta}$.
Analogously, in electrodynamics, we define the Maxwell spinor to be
\[
\phi_{\alpha\beta}(x) = \sigma^{\mu\nu}_{\alpha\beta} F^{\mu\nu}(x) \,.
\]

Let us now turn to the computation of Maxwell and Weyl spinors from the perspective of amplitudes. 
Once again, we discuss the electromagnetic case in more detail. We label our coordinates as $(t^1, t^2, x^1, x^2)$.
The situation involves a single electric charge $Q$ with mass $m$ in a spatially localised wavepacket so that the initial state is
\[
\ket{\psi} = \int \dPhi(p) \, \varphi(p) \, \ket{p}\,.
\]
Recall that the measure $\dPhi(p)$ involves a Heaviside theta function, which imposes positivity of the particle energy. 
However, in split signature there are \emph{two} energy directions: we choose the theta function to impose energy positivity along $t_2$.
On the other hand, we choose to quantise the electromagnetic field so that the annihilation operators annihilate the vacuum at large negative $t_1$.
Our $S$ matrix acting on photon creation and annihilation operators will then be the time evolution operator from large negative $t_1$ to large positive $t_1$.
It is this interplay between boundary conditions which leads to novel phenomena in split signature: as we will see,
this interplay allows for non-trivial solutions of the 
on-shell conditions for the three-point amplitude
\footnote{The boundary conditions were chosen precisely
so that the three-point amplitude would have non-trivial support.}.

We may compute the expectation value of the field strength at large positive $t_1$ using equation~\eqref{eq:Fexpect}. Working in terms of spinors, we find
\[
\phi_{\alpha\beta}(x) = 2 \sqrt{2} \Re \int \dPhi(k) \, i \ket{k}_\alpha \ket{k}_\beta \, e^{-i k \cdot x} \int \dPhi(p',p) \varphi^*(p') \varphi(p) \bra{p'} S^\dagger a_+(k) S\ket{p} \,.
\]
In the classical case, $\varphi^*(p') \varphi(p) \simeq |\varphi(p)|^2$. Working at lowest order in perturbation theory, we may replace 
\[
\bra{p'} S^\dagger a_+(k) S \ket{p} \simeq i \bra{p',k^+} T \ket{p} \,,
\]
so that the Maxwell spinor is
\[
\phi_{\alpha\beta}(x) =- 2 \sqrt{2} \Re \int \dPhi(k) \, \del(2 p \cdot k)\,  \ket{k}_\alpha \ket{k}_\beta \, e^{-i k \cdot x} \, \mathcal{A}(p \rightarrow p' k^+) \,.
\label{eq:MaxwellSpinor}
\]
To arrive at this expression, we performed the integral over the wavefunction, 
taking it to be so sharply peaked that we can set the momentum of 
the incoming particle to its classical expectation value. 
Notice that all the information about the specific interaction 
between our point source and the electromagnetic field is contained 
in the scattering amplitude.
The simplest application of the general formula is to the Coulomb 
case --- then the amplitude, describing a scalar charge interacting 
with a photon, is simply
\[
\mathcal{A}^\textrm{Coul}(p \rightarrow p' k^\pm) = -2 Q p \cdot \varepsilon_\pm(k) \,.
\label{eq:ampCoulomb}
\]
Inserting this amplitude into \eqref{eq:MaxwellSpinor} yields the Maxwell spinor of the Coulomb solution analytically continued to $(2,2)$ signature.

Our justification for the Maxwell spinor~\eqref{eq:MaxwellSpinor} was based on expanding the $S$ matrix, truncating at lowest order in perturbation theory.
However, the result is valid to all orders. This occurs because the out state $S\ket{\psi}$ is actually a coherent state of the electromagnetic field~\cite{Monteiro:2020plf}. The expectation of the field strength operator on this coherent state is equal to its leading perturbative approximation.

From the perspective of scattering amplitudes, equation~\eqref{eq:MaxwellSpinor} is very natural. The amplitude $\mathcal{A}(p \rightarrow p' k^+)$ carries helicity weight which is cancelled by the spinors in the equation. Aside from this, the Maxwell spinor is nothing but an on-shell Fourier transform of the amplitude.
It is the simplest it could be. The natural gravitational equivalent is
\[
\Psi_{\alpha\beta\gamma\delta}(x) = 2 \kappa \Re \int \dPhi(k) \, \del(2 p \cdot k)\,  \ket{k}_\alpha \ket{k}_\beta\ket{k}_\gamma \ket{k}_\delta \, e^{-i k \cdot x} \, i \mathcal{M}(p \rightarrow p' k^+) \,.
\label{eq:WeylSpinor}
\]
This expression is indeed the expectation value of the Weyl spinor as a detailed calculation~\cite{Monteiro:2020plf} demonstrates (this calculation also supplies the overall factor required to agree with the classical normalisation). 
It is worth emphasising again that the outgoing gravitational state $S \ket{\psi}$ is coherent to all orders in perturbation theory.
However, equation~\eqref{eq:WeylSpinor} only computes the 
expectation value of the linearised Riemann tensor operator, which 
is the operator we started with in equation~\eqref{eq:Riemannop}.

The Maxwell spinor~\eqref{eq:MaxwellSpinor} and the Weyl 
spinor~\eqref{eq:WeylSpinor} are very interesting from the 
perspective of
the double copy.
At the level of three-point amplitudes, the double copy from 
electrodynamics (or, equivalently at this order, from Yang--Mills 
theory) to gravity is simple: up to overall factors, the 
gravitational amplitude is the square of the YM amplitude.
The simplest case is the gravitational three-point amplitude 
$\mathcal{M}(p \rightarrow p' k^+)$ involving
a massive particle and a graviton: this is the square of the 
Coulombic amplitude~\eqref{eq:ampCoulomb}.
In this case the Weyl spinor~\eqref{eq:WeylSpinor} is nothing but the Weyl spinor of the linearised Schwarzschild solution.
Therefore the double copy extends beyond the domain of scattering amplitudes and has concrete implications in classical field theory, even for stationary situations.
We will explore this topic in greater detail in the next section of the review.

More generally, equations~\eqref{eq:MaxwellSpinor} and~\eqref{eq:WeylSpinor} present the spinorial curvatures, in on-shell momentum space,
as double copies whenever the relevant scattering amplitudes are double copies of one another. 
Recent progress~\cite{Arkani-Hamed:2017jhn,Guevara:2018wpp,Huang:2019cja,Arkani-Hamed:2019ymq,Guevara:2019fsj,Emond:2020lwi} in our understanding of massive scattering amplitudes in four dimensions has revealed two very simple deformations
of the `Coulomb amplitude'~\eqref{eq:ampCoulomb} and its double copy. 
One deformation\footnote{We present the deformations in Minkowski space.} is~\cite{Huang:2019cja},
\[
\mathcal{A}^\textrm{dyon}(p \rightarrow p' k^\pm) = e^{\pm i \theta} \mathcal{A}^\textrm{Coul}(p \rightarrow p' k^\pm) \,.
\]
The deformation can be easily interpreted by computing the corresponding Maxwell spinor. 
In general, the (Minkowski signature) Maxwell spinor has the structure
\[
\phi_{\alpha \beta} \sim (\v{E} + i \v{B}) \cdot \v{\sigma}_{\alpha\beta} \,,
\label{eq:MaxwellEB}
\]
in terms of electric and magnetic fields. 
Thus the deformation upgrades the simple Coulomb charge to a dyon, with electric and magnetic charges $Q \cos \theta$ and $Q \sin \theta$.
The angle $\theta$ is associated with electric-magnetic duality.
The double copy of this amplitude in gravity leads to the linearised Weyl spinor of the Taub--NUT solution: the NUT parameter is determined by the mass and $\sin \theta$.
Thus we see that NUT charge is the double copy of magnetic charge. This is consistent with earlier findings from the classical double copy \cite{Luna:2015paa}.

The second deformation induces a large, classical spin~\cite{Guevara:2018wpp,Arkani-Hamed:2019ymq}. The deformation is
\[
\mathcal{A}^{\sqrt{\textrm{Kerr}}}(p \rightarrow p' k^\pm) = e^{\mp k \cdot a} \mathcal{A}^\textrm{Coul}(p \rightarrow p' k^\pm) \,.
\]
The parameter $a$ is a 4-vector satisfying $p \cdot a = 0$. 
Referring back to equation~\eqref{eq:MaxwellSpinor}, we find
\[
\phi^{\sqrt{\textrm{Kerr}}}_{\alpha\beta}(x) =- 2 \sqrt{2} \Re \int \dPhi(k) \, \del(2 p \cdot k)\,  \ket{k}_\alpha \ket{k}_\beta \, e^{-i k \cdot (x-ia)} \, \mathcal{A}(p \rightarrow p' k^+) \,.
\label{eq:RootKerrSpinor}
\]
Notice that this spinor satisfies
\[
\phi^{\sqrt{\textrm{Kerr}}}_{\alpha\beta}(x) = \phi^{{\textrm{Coul}}}_{\alpha\beta}(x-i a) \,.
\]
This is the Newman--Janis shift~\cite{Newman:1965tw}, appearing in electrodynamics. For this reason the solution has been termed $\sqrt{\textrm{Kerr}}$. 
Under the double copy, it is related to the Kerr solution, obtained directly via the same Newman--Janis shift from the Schwarzschild solution~\cite{Arkani-Hamed:2019ymq,Guevara:2020xjx}. Again, this is consistent with earlier findings from the classical double copy \cite{Monteiro:2014cda}.
Considering both deformations together~\cite{Emond:2020lwi}, it is easy to see that the linearisation of the Kerr--Taub--NUT solution is a double copy of a spinning dyon solution, obtained via a Newman--Janis shift of the dyon.
It is further possible to consider double copies involving the product of amplitudes for opposite helicities, thereby making contact with solutions involving non-trivial dilatons and axions~\cite{Monteiro:2021ztt}, as we will discuss in the next section.

From the perspective of scattering amplitudes, spacetimes of $(2,2)$ signature are very natural for the study of stationary solutions. 
It is exciting that this spacetime signature has also risen to prominence recently with a different motivation: celestial holography~\cite{Atanasov:2021oyu,Crawley:2021auj,Guevara:2021yud}.
Clearly there is much more to understand in this context.

\section{The Classical Double Copy}
\label{sec:classicaldc}

In earlier sections, we reviewed above direct connections between scattering amplitudes and classical observables. The structural similarity between the gauge theory and gravity examples we considered is clear. This begs the question of how the famed double copy between scattering amplitudes in these two classes of theories manifests itself in classical physics. Previously, we gave examples of how the double copy of amplitudes had implications for classical solutions. In this section, we will take this one step further by formulating the double copy directly in terms of the classical solutions.

\subsection{From amplitudes to classical solutions}

The double copy for scattering amplitudes is discussed in a companion article \cite{SAGEX-22-03}; for a more detailed review, 
see ref.~\cite{Bern:2019prr}. Its schematic form is
\[
\label{eq:KLT}
{\mathcal M} = {\mathcal A} \cdot {\mathcal S}^{-1} \cdot \tilde {\mathcal A}\,,
\]
where ${\mathcal M}$ is a gravity amplitude, ${\mathcal A}$ is a gauge theory amplitude, and ${\mathcal S}$ is an amplitude in the `bi-adjoint' cubic scalar theory. For each scattering amplitude, we have the same number of external particles, with the same momenta. The external states in the gravity amplitude are tensor products of the ones in the gauge theory amplitudes, e.g.~$\varepsilon_{\mu\nu}=\varepsilon_{\mu}\tilde\varepsilon_{\nu}$, with $\varepsilon_{\mu}$ from ${\mathcal A}$ and $\tilde\varepsilon_{\mu}$ from $\tilde{\mathcal A}$. One realisation of the schematic form \eqref{eq:KLT} is to interpret ${\mathcal A}$ as a vector of colour-ordered amplitudes. Then, in the bi-adjoint scalar theory, ${\mathcal S}$ is interpreted as a matrix of bi-colour-ordered amplitudes, whose matrix inverse is ${\mathcal S}^{-1}$. We have suppressed coupling constants in the three types of theory. The realisation of equation~\eqref{eq:KLT} obtained in this way --- i.e.~in terms of colour ordering --- is 
the celebrated KLT relation~\cite{Kawai:1985xq}
in the field-theory limit%
\iffalse, and its origins lie in string theory, namely in the relationship between the scattering of closed strings and open strings\fi. \iffalse Here, we will be interested only in the application to field theory, where\fi
In this limit, it expresses the notion that 
\textit{gravity is a double copy of gauge theory\/}. 
An alternative realisation of equation~\eqref{eq:KLT} is the BCJ double copy \cite{Bern:2008qj,Bern:2010ue}, based on a representation of the amplitudes in terms of trivalent diagrams. In this version, we have
\[
{\mathcal S} = \sum_\Gamma \frac{c_\Gamma\,\tilde c_\Gamma}{\prod_{\alpha_\Gamma}p_{\alpha_\Gamma}^2}\,, \quad\;\;
{\mathcal A} = \sum_\Gamma \frac{n_\Gamma\, c_\Gamma}{\prod_{\alpha_\Gamma}p_{\alpha_\Gamma}^2}\,, \quad \;\;
{\mathcal M} = \sum_\Gamma \frac{n_\Gamma\,\tilde n_\Gamma}{\prod_{\alpha_\Gamma}p_{\alpha_\Gamma}^2}\,,
\label{eq:BCJ}
\]
where the sum is over all trivalent diagrams, the denominators encode the propagator factors, and we have again suppressed couplings. The Lie algebra colour factor $c_\Gamma$ associated to a trivalent diagram is built from the structure constants $f^{abc}$, while the kinematic numerator $n_\Gamma$ depends on the momenta and polarisations of the external particles. In gauge theory, the fact that the colour factors are not all linearly independent (due to Jacobi relations) implies that the choice of kinematic numerators is not unique. The expression for the gravity amplitude in equation~\eqref{eq:BCJ}, with two copies of gauge theory kinematic numerators, is valid if at least one set, say $n_\Gamma$, satisfies the same algebraic identities as a set of colour factors $c_\Gamma$, in particular the Jacobi relations:
\[
c_\Gamma\pm c_{\Gamma'}\pm c_{\Gamma''}=0 \quad \leftrightarrow \quad
n_\Gamma\pm n_{\Gamma'}\pm n_{\Gamma''}=0\,.
\]
Kinematic numerators satisfying this property, known as the {\it colour-kinematics duality}, are called BCJ numerators. For the expressions above, we focused on tree-level amplitudes, but there are analogous loop-level expressions valid for loop integrands.

As we have argued in this review, the double copy is not just an elegant property connecting the $S$-matrices of gauge theory and gravity. It is also an extremely useful tool for practical computations of gravity amplitudes. The range of applications is remarkable: supergravity and superstring amplitudes, effective field theories (for which analogues of the double copy exist) and, for what concerns us in this article, the study of classical general relativity. We have seen that gravitational observables can be constructed from gravitational scattering amplitudes, and these in turn can be obtained with double copy techniques. Note that the relevant amplitudes involve not just gravitons but also massive states, describing black holes or other astrophysical objects of interest, such as neutron stars; in contrast, most older applications of the double copy dealt only with massless states. A crucial aspect of the formalism we reviewed is that it is on-shell, which means that a double copy prescription for scattering amplitudes can be directly applied to classical gravity. `Off-shell' double copy approaches have also been pursued, e.g.~based on worldline effective field theory, building on the successful history of this type of formalism in classical gravity. A parallel line of work has aimed to translate the double copy into the traditional geometric setting of general relativity. Surprisingly, this has provided a complete double-copy interpretation of the most important {\it exact} solutions to the Einstein equations, such as the Schwarzschild and Kerr solutions. In the remainder of this section, we will review various double copy approaches, with emphasis on how the exact solutions story fits in with the double copy of amplitudes.

With hindsight, and to maintain the flow from the previous sections, 
we find it may benefit the reader to present the developments in 
non-chronological order. For instance, the Kerr--Schild double copy 
introduced in ref.~\cite{Monteiro:2014cda} and the convolutional 
double copy first discussed in ref.~\cite{Anastasiou:2014qba} are 
put on a clearer footing by discussing first the on-shell connection 
between solutions and scattering amplitudes.

In the previous sections, we reviewed the connection between the Riemann curvature of spacetime and amplitudes. This direct connection applies to weak fields (the far field in the case of gravitational waves). Indeed, while scattering amplitudes are gauge invariant, in gravity there are no local gauge-invariant observables, except in the weak field approximation because the linearised curvature tensor is indeed gauge invariant. Similarly, the field strength is gauge invariant in linearised Yang--Mills theory, i.e.~electromagnetism. In this context, it is simple to proceed with the derivation that the linearised Schwarzschild solution is a double copy of the Coulomb solution, because the three-point amplitudes that generate the linearised solutions satisfy the double copy. That is, we can take the expressions \eqref{eq:MaxwellSpinor} and \eqref{eq:WeylSpinor} obtained in ref.~\cite{Monteiro:2021ztt}, and recognise that $\mathcal{M}(p \rightarrow p' k^\pm)$ is proportional to $\big( \mathcal{A}(p \rightarrow p' k^\pm) \big)^2$. The proportionality factor has a natural physical interpretation in light of equation~\eqref{eq:KLT}, namely that
\[
\label{eq:MAAS}
\mathcal{M}(p \rightarrow p' k^\pm) = \frac{\big( \mathcal{A}(p \rightarrow p' k^\pm) \big)^2}{\mathcal{S}(p \rightarrow p' k^\pm)} \,,
\]
where we ignore coupling constants and normalisation conventions. Via this three-point KLT relation, we can obtain an amplitude for emission of a massless scalar, which then allows us to construct a solution to the scalar wave equation as,
\[
S(x) = \Re \int \dPhi(k) \, \del(2 p \cdot k)\,  e^{-i k \cdot x} \, \mathcal{S}(p \rightarrow p' k^+) \,.
\label{eq:ScalarSpinor}
\]
This should be interpreted as a linearised solution to the bi-adjoint scalar theory. The scalar solution associated to the Coulomb and Schwarzschild cases, for which $\mathcal{S}$ is just a constant, is simply a (2,2) signature version of the solution $1/r$ in Lorentzian signature sourced by a static point particle, as may be expected. So spin-0, spin-1 and spin-2 all fit into this framework. (It is a triviality to construct analogous solutions for higher-spin linear equations of motion.) In the previous section, we saw how the self-dual part of the field strength/curvature was associated to the amplitude to emit a positive helicity photon/graviton. Likewise, $S(x)$ represents a `self-dual' part of the scalar field. The source of the solution, specified by the scalar amplitudes for positive and negative helicities, may be such that $S\neq\bar S$, so that the full solution is $S+\bar S$, in the same way that the field strength/curvature is built from the self-dual and anti-self-dual parts.

\subsection{Perturbative versus exact double copy}

A natural question is: how does the formula \eqref{eq:MAAS} for three-point amplitudes translate directly into a relation for the linearised classical solutions \eqref{eq:ScalarSpinor}, \eqref{eq:MaxwellSpinor} and \eqref{eq:WeylSpinor}? The latter is a momentum-space expression, whose products turn into convolutions in position space. This leads to the expected answer
\[
\label{eq:WeylDCconv}
\Psi_{\alpha\beta\gamma\delta} = \Phi_{(\alpha\beta} \circ S^{-1} \circ \Phi_{\gamma\delta)}\,.
\]
Here, $\circ$ denotes a convolution, such that $(f\circ g)(x)=\int d^Dy f(x-y)g(y)$, and $S^{-1}(x)$ is formally defined as the inverse of $S(x)$ with respect to the convolution. The idea of a convolutional double copy applying to linearised fields was first explored in ref.~\cite{Anastasiou:2014qba} and further developed in works to be discussed below. For now, let us note that the expression \eqref{eq:WeylDCconv} is not straightforward: its explicit application at linearised level is already intricate, and there have been few attempts to extend it to higher order in perturbation theory.

And yet, one could have hoped that at least some known solutions in general relativity, such as the Schwarzschild and Kerr solutions, possess a double copy description that is both simple and exact (i.e.~not just at linearised level). Remarkably, this is indeed the case. The expression of this fact that most closely resembles equation~\eqref{eq:WeylDCconv} is the Weyl double copy \cite{Luna:2018dpt,Godazgar:2020zbv}: there is a class of exact vacuum gravity solutions whose Weyl spinor obeys the decomposition,
\[
\label{eq:WeylDC}
\Psi_{\alpha\beta\gamma\delta} = \frac{\Phi_{(\alpha\beta}\, \Phi_{\gamma\delta)}}{S} \,,
\]
where the field strength on the right-hand side satisfies the flat-spacetime vacuum Maxwell equations, and the scalar field satisfies the flat-spacetime wave equation. Writing $\Phi_{\alpha\beta}=a_{(\alpha} b_{\beta)}$, we have $\Psi_{\alpha\beta\gamma\delta}\propto a_{(\alpha} b_{\beta} a_\gamma b_{\delta)}$. According to the Petrov classification of the algebraic structure of the Weyl tensor (see e.g.~\cite{Penrose:1960eq}), it is clear that such spacetimes are algebraically special, either of type D ($a_\alpha \not\propto b_\alpha$) or of type N ($a_\alpha \propto b_\alpha$). Type D solutions can be thought of as `Coulombic' gravity solutions, e.g.~the Kerr--Taub--NUT family that includes Schwarzschild. Type N solutions are purely radiative, and the simplest example is that of pp-waves. All type D or N spacetimes admit the form \eqref{eq:WeylDC}. For type D spacetimes (and also pp-waves), this is guaranteed by the existence of a Killing 2-spinor \cite{Walker:1970un}, from which the various objects can be built. In order for the Maxwell and scalar fields on the right-hand side to satisfy flat-spacetime equations of motion, there needs to be a natural map between the curved spacetime and a flat spacetime. In the type D case, this condition is guaranteed by the fact that all such spacetimes admit a Kerr--Schild or double-Kerr--Schild form (explained later) \cite{Plebanski:1976gy}, whereas in the type N case, this condition is valid at least for non-twisting solutions \cite{Godazgar:2020zbv}. In these cases, equation~\eqref{eq:WeylDC} provides a simple and exact double-copy interpretation of the spacetime, as elegant as one could hope for. Notice that it is highly non-trivial that the same solutions can satisfy equation~\eqref{eq:WeylDC} and, at linearised level, equation~\eqref{eq:WeylDCconv}, which is the statement most directly connected to the double copy of three-point amplitudes. This fact relies on the details of the solutions, as worked out recently for Kerr--Taub--NUT spacetimes~\cite{Monteiro:2021ztt}. We will return to this point. More generally, this indicates that we cannot hope for a double copy description of generic dynamical solutions, such as a spacetime representing the scattering of black holes, to be as simple as equation~\eqref{eq:WeylDC}.

The expressions \eqref{eq:WeylDC} and \eqref{eq:WeylDCconv} exemplify the two distinct avenues that have been taken regarding the double copy of classical solutions: either one considers simple formulations valid for exact solutions, which necessarily possess a lot of symmetry (otherwise we couldn't construct them exactly!), or one considers a more general but messier formulation valid for any perturbative solution, starting at linearised level.

Let us discuss first the known instances of an exact double copy. We have already mentioned the Weyl double copy \eqref{eq:WeylDC}. It was introduced in ref.~\cite{Luna:2018dpt} as a four-dimensional curvature-based version of the original exact prescription: the Kerr--Schild double copy  \cite{Monteiro:2014cda}, which we now discuss. Kerr--Schild spacetimes are defined by the property that, for a special choice of coordinates, the metric can be written as an exact deviation from the flat case:
\[
\label{eq:KS}
g_{\mu\nu}=\eta_{\mu\nu} + \phi\,k_\mu\,k_\nu\,,
\]
where $k_\mu$ is null and geodesic with respect to $\eta_{\mu\nu}$, and therefore also with respect to $g_{\mu\nu}$. We have\, $g^{\mu\nu}=\eta^{\mu\nu} - \phi\,k^\mu\,k^\nu$\,, where $k^\mu=\eta^{\mu\nu} k_\nu=g^{\mu\nu} k_\nu$. The crucial simplification is that the Ricci tensor, as a ${1 \choose 1}$-tensor, is linear in the metric deviation:
\[
\label{eq:KSRicci}
R^\mu{}_\nu=\frac{1}{2}  \partial_\alpha \left[\partial^\mu (\phi k^\alpha k_\nu)+\partial_\nu ( \phi k^\alpha k^\mu)-\partial^\alpha ( \phi k^\mu k_\nu)\right]\,,
\]
where we denote $\partial^\mu=\eta^{\mu\nu} \partial_\nu$. This means that the Einstein equations are linear in the `graviton field'. We now restrict further to solutions that are stationary in the sense of invariance with respect to $\partial_t$, where $t$ is a Minkowskian time coordinate of $\eta_{\mu\nu}$, and additionally choose to set $k_t=1$, which can be achieved by rescaling $\phi$. Then, we conclude that
\[
R^{\,t}{}_\mu = \frac{1}{2}\,{ \partial^\nu F_{\mu\nu}}  \,,
\quad \text{for} \quad F_{\mu\nu}=\partial_\mu A_\nu-\partial_\nu A_\mu\,, \quad  A_\mu = \phi \, k_\mu\,.
\]
Hence, any such vacuum gravity solution is associated to a flat-spacetime Maxwell solution. In fact, there is also a stationary solution to the flat-spacetime scalar equation $\,\Box \phi=0\,$, since $R^{\,t}{}_t =\frac1{2}\,\partial^i\partial_i \phi$\,. We have, therefore,
\begin{equation}
\label{eq:dszcopy}
\text{double copy:} \;\;g_{\mu\nu}=\eta_{\mu\nu} + \phi\,k_\mu\,k_\nu\,,
\quad \text{single copy:} \;\;A_\mu = \phi \, k_\mu\,,
\quad \text{zero-th copy:} \;\;\phi\,.
\end{equation}
The simplest example, where $\,\phi=\text{constant}/|\v x|\,$ and $\,k_\mu= (1 , \v x / |\v x|)\,$, gives: the Schwarzschild solution in gravity, where $t$ is not the usual Schwarzschild time coordinate; the Coulomb solution in gauge theory, in a particular gauge; and the scalar solution $\propto 1/r$. That is, it gives exact spherically-symmetric solutions with a static `point source'. This double copy of classical solutions matches the expectation based on scattering amplitudes discussed above. The interesting feature is that here we have exact solutions, because Schwarzschild admits Kerr--Schild coordinates. (Obviously, any Maxwell and scalar wave equation solutions are also exact abelian solutions in Yang--Mills theory and bi-adjoint scalar theory, respectively.) This basic example can be generalised to the full Kerr--Taub--NUT family in vacuum gravity \cite{Luna:2015paa}, where the NUT parameter requires an extension from equation~\eqref{eq:KS} to the double-Kerr--Schild case, i.e.~there are two mutually orthogonal `deviation terms'. The Taub--NUT solution, for instance, is the double copy of a dyon. This is a simple example of how the classical double copy provides an exact map between solutions that were long thought to be analogous. It also agrees with the scattering-amplitudes story discussed above. While the Kerr--Schild double copy does not include all type D cases covered by the Weyl double copy (e.g.~the C-metric, which has no equivalent to the stationarity condition), it has the advantage that the higher-dimensional extension is straightforward. It can also be extended to include a cosmological constant  \cite{Luna:2015paa,Carrillo-Gonzalez:2017iyj,Bahjat-Abbas:2017htu}; see also ref.~\cite{Prabhu:2020avf}.

Notice that, strictly speaking, the double copy 
in equation~\eqref{eq:dszcopy} is just $\phi\,k_\mu\,k_\nu$. It is the `graviton field' that is a double copy, not the full metric. An event horizon, when it exists, is the result of the interplay between $\eta_{\mu\nu}$ and $\phi\,k_\mu\,k_\nu$, so there is no counterpart of it in the single and zero-th copy. We are discussing here vacuum solutions, and not necessarily considering their global structure. For instance, the Kerr solution can represent a black hole or a naked singularity, depending on whether the mass and the rotation parameter satisfy the extremality bound, but here the mass is just a constant prefactor in $\phi$.

Before proceeding, let us address an obvious question. Gravity is meant to be the double copy of Yang--Mils theory, which is also non-linear. In the examples above, however, the single copy is a Maxwell solution, which is indeed a solution of Yang--Mills theory, but where the colour dependence is trivial. From the perspective of the Kerr--Schild double copy, the solution to this puzzle is clear: the exact linearisation of the Einstein equations found in \eqref{eq:KSRicci} is a gravitational counterpart of the linearisation of the Yang--Mills equations, so it makes sense that the single copy is an abelian-type solution. Any solution to the Einstein equations that is known exactly has a large symmetry, whether it is manifest or hidden. It belongs to an integrable sector of the space of solutions, and it is perhaps not surprising that it is `linear' in some sense. This is widely expected not to be the case with dynamical solutions representing, say, the scattering of two black holes. In such scenarios, the non-abelian structure is essential. Exact non-abelian solutions have been considered in the context of the double copy, but again these are special. For instance, it was argued in ref.~\cite{Bahjat-Abbas:2020cyb} that both the abelian Dirac monopole and the non-abelian Wu--Yang monopole can be interpreted as the single copy of the Taub--NUT solution; indeed, they are related by a (singular) gauge transformation. A double-copy interpretation of topological aspects of these solutions, expressed in terms of Wilson lines, was discussed in ref.~\cite{Alfonsi:2020lub}. Non-abelian solutions of the bi-adjoint scalar field have also been explored, with the hope that they will provide an insight into the double copy at non-perturbative level \cite{White:2016jzc,DeSmet:2017rve,Bahjat-Abbas:2018vgo}.

Returning to the Kerr--Schild double copy, we can relate it directly to the earlier linearised-level discussion. For a concrete illustration, we focus on the solutions with rotation generated by the amplitudes (ignoring constant prefactors)
\[
\mathcal{M}(p \rightarrow p' k^\pm) = (p\cdot \varepsilon_{\pm})^2 \,e^{\mp k\cdot a}\,,
\qquad
\mathcal{A}(p \rightarrow p' k^\pm) = (p\cdot \varepsilon_{\pm}) \,e^{\mp k\cdot a}\,.
\]
From these amplitudes, respectively, we get the Weyl spinor of the Kerr solution via equation~\eqref{eq:WeylSpinor}, and the Maxwell spinor of the $\sqrt{\text{Kerr}}$ solution via equation~\eqref{eq:MaxwellSpinor}. This is precisely reproduced by the Kerr--Schild double copy: in equation~\eqref{eq:dszcopy}, the Kerr solution is the double copy, and $\sqrt{\text{Kerr}}$ is the single copy. What about the zero-th copy? The KLT relation \eqref{eq:MAAS} defines $\,\mathcal{S}(p \rightarrow p' k^\pm) = e^{\mp k\cdot a}\,$, and therefore defines also a scalar solution from equation~\eqref{eq:ScalarSpinor}. This solution $S$, which is complex in Lorentzian signature, is the one appearing in the Weyl double copy \eqref{eq:WeylDC} for Kerr, and $\phi = S+\bar S$ is the zero-th copy scalar solution appearing in equation~\eqref{eq:dszcopy} for Kerr. We see that the Kerr--Schild double copy, the Weyl double copy and the amplitudes double copy are all consistent. Notice that there is some freedom in defining the double-copy interpretation of the Kerr solution, because of the freedom to choose the scalar field. In this discussion, we took the amplitudes double copy to be $\mathcal{M}_\pm(a)=\big(\mathcal{A}_\pm(a)\big)^2/\mathcal{S}_\pm(a)$, where $a$ is rotation parameter, and we denoted $\mathcal{S}_\pm(a)=e^{\mp k\cdot a}$, etc. Given the form of the amplitudes, however, we could just as well have taken the double copy to be $\mathcal{M}_\pm(a)=\big(\mathcal{A}_\pm(a/2)\big)^2/\mathcal{S}(0)$. The Kerr--Schild and Weyl prescriptions select the former choice, which is more attuned to the algebraic structure of the solutions, and which makes it possible to turn the statement of the classical double copy from a linearised one into an exact statement. The freedom to choose the scalar, altering what we mean by the zero-th copy and therefore also the single copy, is a generic feature of the classical double copy. In this example, it is related to the fact that the Kerr `particle' is a massive extended object of size $|a|$.\footnote{It may be puzzling that the amplitude for emitting a massless scalar may depend on a $\pm$ polarisation-type sign. The reason is precisely that the massive `particle' is actually an extended object of size $|a|$ in the rotating case.} We have some freedom then to choose whether the scalar is also sourced by an extended object or by a point particle, the latter being the case $S(x)\propto 1/r$ obtained from $\mathcal{S}(0)$. An analogous example would be to consider, instead of a rotation parameter, an electric-magnetic-type duality angle, which can also apply to the scalar solution. For more details, see ref.~\cite{Monteiro:2021ztt}.  In the case of type N solutions, there is even greater freedom in choosing the scalar, as discussed in ref.~\cite{Godazgar:2020zbv}. This is unfamiliar in the older context of scattering amplitudes only because the linearised solutions considered there are straightforward plane waves, whereas the classical double copy admits more complicated solutions. As long as they arise from scattering amplitudes, at least, their grounding in the basic statement \eqref{eq:MAAS} is clear.

Given the difficulty in constructing exact solutions to the Einstein equations, it is not surprising that the exact classical double copy has received strong attention. It motivates a new understanding of the solution space in gravity. The algebraically special solutions covered by the Kerr--Schild and Weyl double copy maps arise in many contexts. One interesting example is the fluid/gravity duality, where the Navier-Stokes equations are seen to arise from a near-horizon expansion in general relativity; see e.g. \cite{Bredberg:2011jq}. Ref.~\cite{Keeler:2020rcv} discussed how the fluid solutions in the fluid/gravity duality relate to Maxwell solutions via the Weyl double copy, finding, for instance, that a constant vorticity fluid maps to a solenoid gauge field. Variations on the Kerr--Schild or Weyl double copy have also been explored. One example includes an accelerating (rather than stationary) source producing Bremsstrahlung radiation, which duly satisfies the double copy between gauge theory and gravity \cite{Luna:2016due}. Another example is a double copy proposal for a class of self-dual metrics, including the Eguchi--Hanson instanton \cite{Berman:2018hwd}; see \cite{Luna:2018dpt} for an alternative discussion of the latter solution. Refs.~\cite{Bah:2019sda,Elor:2020nqe,Farnsworth:2021wvs} explored complexified solutions. Ref.~\cite{Alawadhi:2020jrv} investigated 
the 
occurrence in solutions known in the literature of expressions 
similar to the Weyl double copy. We will later mention other 
examples.

In recent years, there has been a renewed interest in the topic of asymptotic symmetries and their role in a celestial holographic description of asymptotically flat spacetimes. The application to gravity and gauge theory is a natural setting for the double copy. The electric-magnetic duality in electromagnetism, interpolating between the Coulomb solution and a magnetic monopole,  was identified in refs.~\cite{Huang:2019cja,Alawadhi:2019urr,Banerjee:2019saj} as the single copy of a BMS supertranslation interpolating between Schwarzschild and pure NUT, known as an Ehlers transformation. This applies to exact solutions, but is easier to understand when looking at the three-point amplitudes that generate the linearised solutions, namely
\begin{align}
\label{eq:Adyon}
    \mathcal{A}^\textrm{dyon}(p \rightarrow p' k^\pm) & = e^{\pm i \theta} \mathcal{A}^\textrm{Coul}(p \rightarrow p' k^\pm) \,, \\
        \mathcal{M}^\textrm{Taub--NUT}(p \rightarrow p' k^\pm) & = e^{\pm i \theta} \mathcal{M}^\textrm{Schw}(p \rightarrow p' k^\pm) \,.
\label{eq:Mtaubnut}
\end{align}
Asymptotic symmetries have also been explored in the non-abelian case, restricting to self-dual gauge fields and metrics \cite{Campiglia:2021srh}. It was shown that the asymptotic symmetry transformations preserving the self-duality conditions in gauge theory and gravity were related by the double copy. The self-dual sectors are optimal settings for the double copy, because the colour-kinematics duality can be made manifest, as shown in ref.~\cite{Monteiro:2011pc}. In particular, there is a Lie algebra of area-preserving diffeomorphisms that is the kinematic counterpart of the colour Lie algebra. See also ref.~\cite{Chacon:2020fmr} for deformations of the self-dual case.

Various works have explored asymptotic realisations of the classical double copy. In ref.~\cite{Pasterski:2020pdk}, conformal primary wavefunctions of spin 2, which are associated to type N spacetimes, were shown to be related by a celestial version of the double copy to conformal primary wavefunctions of spin 1. This work was partly motivated by the realisation of the double copy for celestial amplitudes \cite{Casali:2020vuy}. In ref.~\cite{Godazgar:2021iae} and \cite{Adamo:2021dfg}, the classical double copy was investigated for the leading behaviour of asymptotically flat solutions around null infinity, turning into a relation between the leading coefficients of that expansion. The goal is to provide a double copy of 
characteristic initial data, which determines the solutions away 
from null infinity. The C-metric, first described as a double copy in ref.~\cite{Luna:2018dpt}, has been interpreted as expressing a finite superrotation in ref.~\cite{Strominger:2016wns}; in 
ref.~\cite{Godazgar:2021iae}, its single copy was interpreted as a large gauge transformation, as expected by the double copy. See also ref.~\cite{Mao:2021kxq} for related work.

Since scattering amplitudes are directly associated to asymptotic states, it is not surprising that they encode the asymptotic information of the solutions that they generate, as in the discussion of \eqref{eq:Adyon} and \eqref{eq:Mtaubnut}. Similar $(2,2)$ signature constructions of linearised solutions from scattering amplitudes to that presented in refs.~\cite{Monteiro:2020plf,Monteiro:2021ztt} were discussed in refs.~\cite{Crawley:2021auj,Guevara:2021yud}. Ref.~\cite{Crawley:2021auj} further studied the global structure of exact Kerr--Taub--NUT solutions in this signature. In ref.~\cite{Guevara:2021yud}, the construction of linearised solutions from scattering amplitudes made use of twistor variables.

Twistors have played an important role in the modern study of scattering amplitudes since the discovery almost two decades ago that certain amplitudes have a simple geometric description in twistor space \cite{Witten:2003nn}; see ref.~\cite{Elvang:2013cua} for an introduction. In the discussion above, we derived the linearised version of the classical double copy from three-point amplitudes in momentum space. There is an alternative derivation from twistor space, introduced in ref.~\cite{White:2020sfn}. Very briefly, twistor space can be identified with $\mathbb{CP}^3$, and we denote a point in twistor space as $Z^A=(\mu^{\dot\alpha},\lambda_\alpha)\in \mathbb{CP}^3$, $A=1,\ldots,4$. The relation between spacetime fields and twistor space representatives is given by the Penrose transform. This is a correspondence between a solution of the linearised equation of motion for a massless field of spin $s$, which has `curvature' spinor $\varphi_{\alpha_1\alpha_2\cdots\alpha_{2s}}(x)$, and a twistor space cohomology class of homogeneity weight $w=-(s+2)$, i.e.~satisfying $f_{[w]}(LZ^A)=L^{w}f_{[w]}(Z^A)$. The Penrose transform is
\[
\varphi_{\alpha_1\alpha_2\cdots\alpha_{2s}}(x) = \frac1{2\pi i} \oint_\Gamma (\lambda_\beta d\lambda^\beta)\, \lambda_{\alpha_1}\lambda_{\alpha_2}\cdots\lambda_{\alpha_{2s}} \; f_{[-s-2]}(\mu^{\dot\alpha},\lambda_\alpha)\big|_{\mu^{\dot\alpha}=x^{\alpha\dot\alpha}\lambda_\alpha} \,,
\]
where $\Gamma$ is a contour in $\mathbb{CP}^1$ with homogeneous coordinates $\lambda_\alpha$. The restriction $\mu^{\dot\alpha}=x^{\alpha\dot\alpha}\lambda_\alpha$ is known as the incidence relation. Ref.~\cite{White:2020sfn} described how the Weyl double copy relation \eqref{eq:WeylDC} follows from the Penrose transform via a twistor space double copy,
\[
\label{eq:twistorDC}
f_{[-6]} = \frac{f_{[-4]}\tilde f_{[-4]}}{f_{[-2]}}\,,
\]
for suitably chosen twistor representatives. In fact, it allows for $f_{[-4]}\neq \tilde f_{[-4]}$, representing a double copy involving two distinct gauge fields (more on this possibility later). It leads to examples beyond type D and type N, namely of type III, although the conclusion is restricted to linearised solutions. One outstanding question is what determines the twistor representatives within the appropriate cohomology class that make equation~\eqref{eq:twistorDC} valid \cite{Chacon:2021wbr,Chacon:2021lox,Guevara:2021yud}. The twistor approach has been used to provide a double-copy interpretation of the multipole expansion of vacuum type-D solutions \cite{Chacon:2021hfe}. An earlier instance of a double copy of multipoles, for describing spinning matter, was given 
in refs.~\cite{Bautista:2019tdr,Bautista:2019evw}.

Finally, regarding exact solutions, an interesting line of related work addresses the questions of how to define scattering amplitudes in curved backgrounds, rather than in Minkowski spacetime, and how the double copy applies to perturbation theory in such backgrounds. Refs.~\cite{Adamo:2017nia,Adamo:2018mpq} studied scattering amplitudes on `sandwich' plane wave backgrounds in both Yang--Mills theory and gravity, finding that the memory effect caused by the plane wave on the scattered states exhibits a double copy. One area where this type of question --- of scattering on strong backgrounds --- is commonly asked is strong field quantum-electrodynamics. Motivated by this connection, aspects of the classical double copy have been investigated 
in refs.~\cite{Ilderton:2018lsf,Andrzejewski:2019hub,Adamo:2020qru}.

\subsection{Beyond vacuum gravity}

So far, we have discussed vacuum gravity solutions obtained as the `square' of a gauge-theory solution, e.g.~Schwarzschild as the square of Coulomb. However, the double copy can involve the product of two distinct gauge-theory solutions. In general, the double copy of Yang--Mills theory is the theory of a massless rank-2 tensor field, i.e.~its perturbative spectrum includes not just the graviton, but also the dilaton and the B-field (anti-symmetric 2-form field). This theory is the universal massless sector of supergravity, and arises from the zero-mass level of the closed string. It is sometimes known as ${\mathcal N}=0$ supergravity or NS-NS gravity. Pure Einstein gravity (vanishing dilaton and B-field) arises as a consistent truncation at the classical level. The double copy of two distinct gauge-theory solutions, and even the most general `square' of a single solution, require us to work with NS-NS gravity.

It is instructive to look at the double copy of linearised plane waves, which are the asymptotic states in scattering amplitudes. Given two gauge-theory plane waves with null momentum $k_\mu$, and polarisations $\varepsilon_\mu$ and $\tilde\varepsilon_\mu$, we can take $\varepsilon_\mu\tilde \varepsilon_\nu$ as a straightforward double copy. And yet, a moment's reflection opens the possibilities of considering $ \varepsilon_{(\mu} \tilde\varepsilon_{\nu)}$ or $ \varepsilon_{[\mu} \tilde\varepsilon_{\nu]}$. There is also the possibility of considering $\varepsilon\cdot\tilde\varepsilon\; \Delta_{\mu\nu}$; here, $\Delta_{\mu\nu}= \eta_{\mu\nu}-(k_\mu q_\nu+k_\nu q_\mu)/(k\cdot q)$ projects into the space where the polarisation vectors live, taken to be orthogonal to $k_\mu$ and to a null reference momentum $q_\mu$ (a gauge choice). The most general double copy of the two gauge theory plane waves is, therefore, the linear combination
\[
\label{eq:polgrav}
\varepsilon_{\mu\nu}=  c^{h}\left(\varepsilon_{(\mu} \tilde\varepsilon_{\nu)}-\frac{\Delta_{\mu\nu}}{D-2} \,\varepsilon\cdot\tilde\varepsilon \right)+  c^{B}\,\varepsilon_{[\mu} \tilde\varepsilon_{\nu]} +  c^{\phi}\,\frac{\Delta_{\mu\nu}}{D-2} \,\varepsilon\cdot\tilde\varepsilon\,,
\]
where we have the graviton, B-field and dilaton contributions. (We work in the Einstein frame, where the propagator does not mix the graviton and the dilaton.)

We can express the double copy of classical solutions at linearised level exactly as in \eqref{eq:polgrav}, but in position space, as discussed in ref.~\cite{Luna:2016hge} based on a `fat graviton' field incorporating graviton, dilaton and B-field. For the `square' of the Coulomb solution, for instance, the Schwarzschild solution arises by picking only the graviton contribution, with the mass parameter associated to $c^h$. There can be no B-field contribution because this requires two distinct gauge-theory solutions, analogously to $\varepsilon_{[\mu} \varepsilon_{\nu]}=0$. However, a dilaton contribution is allowed by turning on $c^{\phi}$. The solution obtained in this way is known exactly: it is the unique static, spherically symmetric and asymptotically flat solution to the Einstein equations with a minimally coupled scalar field, discovered by Newman, Janis and Winicour (JNW) \cite{Janis:1968zz}. This two-parameter family ($c^{h},c^{\phi}$) is the most general double copy of the Coulomb solution. The Schwarzschild solution is the special case of vacuum gravity, $c^{\phi}=0$; all the other cases have a naked singularity at the origin (cf. uniqueness theorems).

Recall that the direct connection between scattering amplitudes and the linearised classical double copy is manifest at the level of `curvatures', i.e.~the curvature tensor in gravity and the field strength in gauge theory. In order to extend this beyond vacuum gravity, ref.~\cite{Monteiro:2021ztt} introduced a generalised curvature, based on Riemann--Cartan geometry, which incorporates the graviton, the dilaton and the B-field. This is precisely the curvature of the fat graviton field $H_{\mu\nu}$ of \cite{Luna:2016hge}. At linearised level, we have ${\mathcal R}_{\mu\nu}{}^{\rho\sigma}= -\frac{\kappa}{2}\,
\partial_{[\mu}\partial^{[\rho}H_{\nu]}{}^{\sigma]}$, which generalises the standard Riemann curvature, related to the ordinary graviton via $R_{\mu\nu}{}^{\rho\sigma}= -\frac{\kappa}{2}\,
\partial_{[\mu}\partial^{[\rho}h_{\nu]}{}^{\sigma]}$; in both instances, the linearised curvatures are gauge invariant, which is why they can relate directly to amplitudes. The connection to amplitudes is easier to express using the spinorial form of the generalised curvature tensor in four dimensions:
\[
\label{eq:gencurvspinors}
{\mathcal R}_{\alpha\dot\alpha\beta\dot\beta\gamma\dot\gamma\delta\dot\delta} =
X_{\alpha\beta\gamma\delta} \,\bar\epsilon_{\dot\alpha\dot\beta}\,\bar\epsilon_{\dot\gamma\dot\delta} + \bar X_{\dot\alpha\dot\beta\dot\gamma\dot\delta}\,\epsilon_{\alpha\beta}\,\epsilon_{\gamma\delta} + 
\Phi_{\alpha\beta\dot\gamma\dot\delta} \,\bar\epsilon_{\dot\alpha\dot\beta}\,\epsilon_{\gamma\delta} + \bar\Phi_{\dot\alpha\dot\beta\gamma\delta}\,\epsilon_{\alpha\beta}\,\bar\epsilon_{\dot\gamma\dot\delta}\,.
\]
At the linearised level, $X_{\alpha\beta\gamma\delta}=\Psi_{\alpha\beta\gamma\delta}$ is the self-dual graviton contribution, i.e.~the Weyl spinor encountered previously. In the connection to 3-pt amplitudes seen in \eqref{eq:WeylSpinor}, this component arises from the polarisation tensor $\varepsilon_\mu^+\varepsilon_\nu^+$. Similarly, $\bar\Psi_{\alpha\beta\gamma\delta}$ arises from $\varepsilon_\mu^-\varepsilon_\nu^-$. The two other cases, $\varepsilon_\mu^+\varepsilon_\nu^-$ and $\varepsilon_\mu^-\varepsilon_\nu^+$, are associated to $\Phi_{\alpha\beta\dot\gamma\dot\delta}$ and $\bar\Phi_{\dot\alpha\dot\beta\gamma\delta}$. These contributions arise from the dilaton and the axion (the single degree of freedom of the B-field in four dimensions).

An elegant interplay of parameters occurs in the components of the generalised curvature when we consider the double copy of rotating dyons, building on our previous discussions. Recalling section~\ref{sec:continuation}, a rotating dyon is generated by the (Lorentzian) 3-pt amplitudes
\[
e^{\pm (-k \cdot a+i\theta)} \mathcal{A}^{\textrm{Coul}}_\pm\,,
\]
where $\mathcal{A}^{\textrm{Coul}}_\pm=\mathcal{A}^\textrm{Coul}(p \rightarrow p' k^\pm)$ generate the Coulomb solution. The self-dual field strength spinor is given by
\[
\phi_{\alpha\beta}(x) &= e^{i\theta}\,\phi^\textrm{Coul}_{\alpha\beta}(x-ia)~.
%\\[1em]
%\bar\phi_{\dot\alpha\dot\beta}(x) &= e^{-i\theta}\,\bar\phi^\textrm{Coul}_{\dot\alpha\dot\beta}(x-ia)~.
\]
For the double copy, we take the product of left (L) and right (R) amplitudes:
\[
e^{\eta_L (-k \cdot a_L+i\theta_L)} \mathcal{A}^{\textrm{Coul}}_{\eta_L}\;
\times \; e^{\eta_R (-k \cdot a_R+i\theta_R)} \mathcal{A}^{\textrm{Coul}}_{\eta_R} \;
\,,
\]
where $\mathcal{A}^{\textrm{Coul}}_{\eta_L}\times\mathcal{A}^{\textrm{Coul}}_{\eta_R}$ generate the JNW solution. There are four cases, according to the signs $\eta_L$ and $\eta_R$, leading to four linearised curvature spinors associated to equation \eqref{eq:gencurvspinors}. These spinors are
\[
{\psi}_{\alpha\beta\gamma\delta}(x)
&=e^{i(\theta_L+\theta_R)}\,{\psi}_{\alpha\beta\gamma\delta}^\text{JNW}\big(x-i(a_L+a_R)\big)~,
%\\[1em]
%\bar{\psi}_{\dot\alpha\dot\beta\dot\gamma\dot\delta}(x)
%&=e^{-i(\theta_L+\theta_R)}\,\tilde{\psi}_{\dot\alpha\dot\beta\dot\gamma\dot\delta}^\text{JNW}\big(x-i(a_L+a_R)\big)~,
\\[1em]
{{\Phi}}_{\alpha\beta\dot\gamma\dot\delta}(x)
&=e^{i(\theta_L-\theta_R)}\,{{\Phi}}_{\alpha\beta\dot\gamma\dot\delta}^\text{JNW}\big(x-i(a_L-a_R)\big)~,
%\\[1em]
%{\tilde{ {\Phi}}}_{\dot\alpha\dot\beta\gamma\delta}(x)
%&=e^{-i(\theta_L-\theta_R)}\,{\tilde{ {\Phi}}}_{\dot\alpha\dot\beta\gamma\delta}^\text{JNW}\big(x-i(a_L-a_R)\big)~.
\]
and their conjugate spinors; see ref.~\cite{Monteiro:2021ztt} for details.
Notice how the parameters $\theta_L+\theta_R$ and $a_L+a_R$ are associated to the graviton spinors, while $\theta_L-\theta_R$ and $a_L-a_R$ are associated to the dilaton/axion spinors. Each of the linearised curvature spinors satisfies the equations of motion independently. Indeed, for the graviton spinor on the right-hand 
side, we could have substituted the superscript `JNW' by `Schwarzschild'. The graviton spinor on 
the left-hand side is that of a linearised Kerr--Taub--NUT solution.  Let us now ignore the rotation and focus on $\theta_L$, $\theta_R$. In gravity, via the double copy, these lead to two distinct parameters: $\theta_L+\theta_R$ corresponds to the Ehlers duality transformation of the Schwarzschild spinors, as discussed above \eqref{eq:Mtaubnut}, while $\theta_L-\theta_R$ corresponds to the supergravity dilaton-axion duality rotation. Both duality transformations in gravity arise as a double copy of electric-magnetic duality in gauge theory! Considering now the rotation parameters, it is interesting that there is a new Newman--Janis-like shift for the axion and the dilaton, at least at linearised level.

We mentioned that the generalised curvature is the curvature of a `fat graviton' incorporating the graviton, the dilaton and the B-field. There is a natural map from the gravity 3-pt amplitudes to the generalised curvature, e.g.~equation\eqref{eq:WeylSpinor}; see ref.~\cite{Monteiro:2021ztt} for more details. This map is straightforward because the linearised curvature is gauge invariant. We can obtain a map directly to the fat graviton (equivalently, to its graviton, dilaton and B-field parts), but in this case one needs to specify a gauge. Working with a gauge-invariant object is certainly more in the spirit of the scattering amplitudes methods. Nevertheless, one may be interested in looking at the fields, and taking them off-shell, in order to construct a Lagrangian, for instance.

In this context, let us revisit the convolutional double copy map, mentioned earlier in equation~\eqref{eq:WeylDCconv}. There should be a suitably defined map that applies generally, at least at linearised level. As observed in ref.~\cite{Anastasiou:2014qba}, the linearised map from Yang--Mills theory to gravity should take the schematic form
\[
H_{\mu\nu} = A^a_\mu \circ \phi^{-1}_{ab} \circ \tilde A^b_\nu\,,
\]
where $\phi^{-1}_{ab}$ is an inverse of the bi-adjoint scalar field with respect to the convolution, and $H_{\mu\nu}$ should be interpreted as what we called the fat graviton. This schematic form is the counterpart of the polarisations relationship $\varepsilon_{\mu\nu}=\varepsilon_\mu\tilde \varepsilon_\nu$. Just as for the latter, the linearised gauge transformations match on both sides, which can be interpreted as the gauge symmetry in gravity arising from that in gauge theory \cite{Anastasiou:2014qba}. The precise definition of the convolutional map is not straightforward. It is easier to apply it to solutions of the linearised equations of motion (i.e.~on-shell), as in 
refs.~\cite{Cardoso:2016ngt,Cardoso:2016amd}. It is trickier, as one may expect, to apply it to off-shell fields, as discussed in refs.~\cite{Anastasiou:2018rdx,LopesCardoso:2018xes}. Ref.~\cite{Anastasiou:2018rdx} described how the off-shell degrees of freedom in gravity are the double copy of those in gauge theory when the ghosts are included, which leads to a double copy map for the BRST quadratic Lagrangians. Ref.~\cite{Luna:2020adi} presented a construction of the linearised JNW solution as a convolutional double copy of the Coulomb solution, with the ghosts playing a role. Refs.~\cite{Borsten:2020xbt,Borsten:2020zgj,Borsten:2021hua} took these insights beyond linear theory with the aim of describing the double copy structure of the interacting Lagrangian. An alternative approach to the matching of the off-shell degrees of freedom and the structure of the interacting Lagrangian has presented in ref.~\cite{Ferrero:2020vww}.
The convolutional map has also been applied with certain non-trivial backgrounds \cite{Borsten:2019prq,Borsten:2021zir}. 

Returning to the comparison between equation~\eqref{eq:WeylDCconv}, based on the convolutional double copy, and equation~\eqref{eq:WeylDC}, valid for certain exact solutions, it may be helpful to see an example of how they can both apply, based on ref.~\cite{Monteiro:2021ztt}. We consider the particularly simple example of the Coulomb solution, which leads via the double copy to the JNW solution. The graviton contribution to the generalised curvature tensor $\mathcal{R}_{\mu\nu\rho\sigma}$ is, up to a constant prefactor,
\[
\label{eq:WeylPFF}
P^{\mu\nu\rho\sigma}_{\tau\lambda\eta\omega}\ \left(F^{\tau\lambda} \circ S^{-1}\circ F^{\eta\omega}\right) = -3\, P^{\mu\nu\rho\sigma}_{\tau\lambda\eta\omega}\,\frac{F^{\tau\lambda}F^{\eta\omega}}{S}\,,
\]
where $S=1/r$\,, and $P^{\mu\nu\rho\sigma}_{\tau\lambda\eta\omega}$ projects out the traces, leading to the Weyl tensor. The non-trivial properties of the solutions and the projector make it possible to disentangle the convolutions \cite{Monteiro:2021ztt}. Then, the expression above is simply a tensorial (rather than spinorial) version of the Weyl double copy. However, the same simplification does not arise in the dilaton contribution to $\mathcal{R}_{\mu\nu\rho\sigma}$, which, up to a constant prefactor, turns out to be
\[
\label{eq:FFSdilaton}
\left( F^{\lambda[\mu}\circ S^{-1}\circ F_{\lambda[\rho}\right){\delta^{\nu]}}_{\sigma]}
= -3\,\frac{F^{\lambda[\mu}\,\delta^{\nu]}_{[\rho}\,F_{\sigma]\lambda}}{S}
	+(4\pi)^2 S^3
	\left(
		\delta^{[\mu}_{[\rho}\delta^{\nu]}_{\sigma]}-
		\delta^{[\mu}_{[\rho}u^{\nu]}\,u_{\sigma]}
	\right)\,.
\]
Due to the last term, a Weyl-type double copy for the dilaton contribution does not take a simple form in position space. A remarkable simplification like equation~\eqref{eq:WeylPFF} is only possible in special cases. Fortunately, some of the most important known exact solutions fall into this category. In fact, for JNW, while the solution with dilaton does not admit a simple position-space Weyl double copy --- at least as currently understood --- it does admit an exact double copy of Kerr--Schild type, as we now discuss.

There is a well-known `doubled' formalism that applies specifically to NS-NS gravity, which is double field theory (DFT) \cite{Hull:2009mi}; see refs.~\cite{Aldazabal:2013sca,Hohm:2013bwa} for an introduction. Very briefly, DFT is an approach to the low-energy limit of closed string theory that makes the action of T-duality manifest, by considering a doubled geometry with points labelled by $X^M=(x^\mu,\tilde x_\nu)$. For a toroidal spacetime in $D$ dimensions, these coordinates on the doubled $2D$-dimensional space are conjugate to momenta and winding. The idea is that T-duality acts linearly on the doubled space, by the action of $O(D,D)$. Therefore, $O(D,D)$-covariance ensures manifest T-duality. For our purposes, what matters is that the NS-NS fields are nicely packaged into a `generalised metric' $\mathcal H_{MN}$, which is an $O(D,D)$ tensor, and a `DFT dilaton', which is an $O(D,D)$ scalar. The connection to the classical double copy was introduced in ref.~\cite{Lee:2018gxc}, via a DFT Kerr--Schild-type ansatz for the generalised metric:
\[
\label{eq:DFTKS}
\mathcal{H}_{MN} = \mathcal{H}_{0 MN} +   \phi \big(K_{M}\bar{K}_{N} + K_{N}\bar{K}_{M}\big)\,,
\]
where $\mathcal{H}_{0 MN}$ is a trivial generalised metric associated to Minkowski spacetime. We won't present details, but hopefully the analogy with the usual Kerr--Schild ansatz \eqref{eq:KS} is clear. $K_{M}$ and $\bar K_{M}$ are built from null vectors $k_\mu$ and $\bar k_\mu$, respectively, in such a way that they are chiral and anti-chiral with respect to a notion of chirality that can be traced back to the chirality on the closed string worldsheet. In this way, they are associated to left- and right-movers on the worldsheet --- the same decomposition that lies at the heart of the KLT relations in string theory \cite{Kawai:1985xq}. Since $k_\mu$ and $\bar k_\mu$ are each associated to a solution to the Maxwell equations, similarly to the usual Kerr--Schild double copy, the `left- and right-moving gauge fields' provide a KLT interpretation of the Kerr--Schild double copy. Ref.~\cite{Lee:2018gxc} provided the first examples of exact double copy involving the dilaton and the B-field. More importantly, it described how the `double' in double copy and double field theory can be identified. The exact double-copy interpretation of the JNW solution follows an ansatz which is a small generalisation of equation~\eqref{eq:DFTKS} \cite{Kim:2019jwm}. The generalisation means that the map is not `local' in position space, because the relation between the gauge fields and the gravity fields involves integrations. This may have some relation to the feature found in equation~\eqref{eq:FFSdilaton}.

There are also examples of the classical double copy beyond NS-NS gravity. So far we only considered pure NS-NS gravity. The introduction of a gauge field in the gravity theory has been investigated thoroughly in the context of scattering amplitudes; see e.g.{} ref.~\cite{Chiodaroli:2014xia}, or see ref.~\cite{Bern:2019prr} for a review. One obtains a `heterotic gravity theory': the gauge field is coupled not only to the graviton, but also to the dilaton and the B-field, in the manner dictated by the zero-mass level of the heterotic string. This perturbative spectrum is reproduced straightforwardly from the product of two copies of Yang--Mills theory, one of which is coupled to a scalar field. For illustration, the tensor product in four dimensions gives,
$$
(\varepsilon^+,\varepsilon^-) \otimes (\varepsilon^+,\varepsilon^-,\varphi)
\;=\; (\varepsilon^+\otimes\varepsilon^+,\varepsilon^-\otimes\varepsilon^-,\varepsilon^+\otimes\varepsilon^-,\varepsilon^-\otimes\varepsilon^+,\varepsilon^+\otimes\varphi,\varepsilon^-\otimes\varphi)\,,
$$
where $\varphi$ is the scalar wavefunction. The NS-NS gravity spectrum is supplemented by a gauge field with its two polarisations. 
Building on this map of states studied for scattering amplitudes, examples of linearised double copy of black holes with gauge fields were presented in refs.~\cite{Cardoso:2016ngt,Cardoso:2016amd}. In ref.~\cite{Cho:2019ype}, a Kerr--Schild-type ansatz in heterotic DFT provided examples of exact double copy, an idea further developed in refs.~\cite{Lescano:2020nve,Lescano:2021ooe,Angus:2021zhy,Lescano:2022nhp}. Extensions to exceptional field theories, which arise in the low-energy limit of string/M-theory, were presented in refs.~\cite{Berman:2020xvs,Berman:2022bgy}.

One important point to keep in mind is that pure Einstein--Yang--Mills theory is not known to admit a double-copy description. In many instances in the amplitudes literature, people refer to Einstein--Yang--Mills theory when they really mean the heterotic theory with dilaton and B-field. The pure Einstein--Yang--Mills theory is not (even classically) a consistent truncation of the heterotic theory such that the dilaton and the B-field are turned off, because a non-vanishing gauge field generically sources these fields. Said in a different way, the tree-level amplitudes involving only external gravitons and gluons do not agree in the pure Einstein--Yang--Mills theory and in the heterotic theory, because the dilaton and B-field participate in the scattering in the latter case. This feature has consequences also for the classical double copy. For instance, the Kerr--Newman black hole is a solution to the pure Einstein--Yang--Mills equations (in particular, to the pure Einstein--Maxwell equations), but is not a solution to the heterotic theory. This is because turning on the `Coulombic' gauge field necessarily backreacts on the dilaton non-linearly. The appropriate black hole solution is the dilatonic charged black hole, whose exact double copy description was given in ref.~\cite{Cho:2019ype}. The fact that the Kerr--Newman spacetime admits Kerr--Schild coordinates and has a Weyl tensor of type D does not imply that it has a double-copy interpretation, at least not in the sense that the double copy is typically understood. The connection between the linearised solution and 3-pt scattering amplitudes, investigated in refs.~\cite{Moynihan:2019bor,Chung:2019yfs}, holds as expected. For a discussion related to this paragraph, see the last appendix of ref.~\cite{Godazgar:2021iae}. Recently, in ref.~\cite{Easson:2021asd}, an extension of the vacuum Weyl double copy was proposed such that the gravity, gauge and scalar fields satisfy equations of motion with sources. It would be interesting to check whether this idea can provide examples with a complete double copy description, which  should now include the Ricci part of the curvature as well as the sources.

Another interesting setting beyond vacuum gravity is that of three-dimensional spacetimes. Gluons have a single physical polarisation in three dimensions, so the double copy of Yang--Mills theory has a single propagating degree of freedom. Indeed, in three dimensions, the only propagating degree of freedom of NS-NS gravity is the dilaton. The double copy of a point charge has been investigated from the perspective of a non-vacuum Kerr--Schild double copy in refs.~\cite{CarrilloGonzalez:2019gof,Alkac:2021seh}. Refs.~\cite{Burger:2021wss,Emond:2021lfy} discussed the application of the double copy to anyons and to topological quantisation conditions. Another interesting example of the double copy relating three-dimensional theories is that of topologically massive gravity and topologically massive Yang--Mills theory  \cite{Moynihan:2020ejh,Gonzalez:2021bes,Moynihan:2021rwh,Hang:2021oso,Gonzalez:2021ztm}. In this context, a natural analogue of the Weyl double copy has recently been presented: the Cotton double copy, based on the Cotton tensor \cite{Emond:2022uaf,Gonzalez:2022otg}.

The classical double copy has also been used to investigate non-singular black holes in string theory and other settings \cite{Pasarin:2020qoa,Easson:2020esh,Mkrtchyan:2022ulc}.

\subsection{Higher-order perturbative double copy}

In the previous subsection, we discussed a large variety of exact classical gravity solutions, in various theories and in various dimensions. And yet, all can be related via the double copy to abelian gauge-theory solutions. This is a manifestation of the powerful symmetries that allow us to write those solutions down explicitly in the first place. The symmetries indicate that the solutions are linear for special choices of (possibly complexified) coordinates. Take the Schwarzschild solution, which is a Kerr--Schild spacetime. Perturbative field theory constructions of the Schwarzschild solution unguided by its symmetries have been, to date, very laborious, and fail to capture the resummation that yields an exact solution \cite{Duff:1973zz,Neill:2013wsa,Mougiakakos:2020laz}. (The Aichelburg--Sexl shockwave solution, which arises from Schwarzschild via an ultra-relativistic boost, is easier to construct, and the perturbative double copy has been verified there to all orders \cite{Saotome:2012vy}; see also ref.~\cite{Cristofoli:2020hnk} for a more recent approach building on the KMOC formalism.) If one considers more-generic spacetimes of interest --- e.g.~those describing the scattering of a pair of black holes, or a black hole binary --- then no special symmetries are known to arise. An exact analytical approach is, therefore, not an option: either one considers a perturbative analytical approach of some type (in this article, we focus on post-Minkowskian), or the only option is numerical relativity.

The double copy has been applied perturbatively to classical gravity in various ways, which fall into two groups: on-shell methods (i.e.~strictly scattering amplitudes) and off-shell methods. Examples of off-shell methods applied beyond linearised order include worldline effective field theory approaches \cite{Goldberger:2016iau,Shen:2018ebu,Goldberger:2017frp,Goldberger:2017vcg,Goldberger:2017ogt,Goldberger:2019xef,Plefka:2018dpa,Plefka:2019hmz,Plefka:2019wyg,Shi:2021qsb} and the `fat graviton' construction mentioned in the previous section \cite{Luna:2016hge,Kim:2019jwm}. In many examples, the relation between off-shell quantities and amplitudes may be enough to understand how to apply the double copy off-shell, but a generic method is still missing. Recall that the linearised off-shell double copy of ref.~\cite{Anastasiou:2018rdx} involved ghosts. Indeed, the double copy is only known to be straightforward when fully off-shell if we restrict to the self-dual sector and use light-cone gauge \cite{Monteiro:2011pc}. Recently, however, progress has been made towards understanding the perturbative double copy structure of classical fields, rather than amplitudes. Ref.~\cite{Cheung:2021zvb} described a `covariant colour-kinematics duality' at the level of the equations of motion. Ref.~\cite{Cho:2021nim} showed that currents (one leg off-shell) obtained from perturbative double field theory satisfy a simple generalisation of the on-shell KLT relations; see \cite{Hohm:2011dz,Diaz-Jaramillo:2021wtl,Bonezzi:2022yuh} for related work. Regarding complete `off-shellness', it is worth mentioning that there exist elegant and useful Lagrangians for gravity that were at least motivated by the double copy \cite{Cheung:2016say,Cheung:2017kzx}.

The application of the double copy to on-shell methods for classical gravity is very natural, as we hope this review article has conveyed. In the important case of the gravitational two-body problem, both the KLT prescription and the BCJ prescription have been used. For instance, ref.~\cite{Bern:2019nnu} and subsequent work used the KLT relation for tree amplitudes appearing in unitarity cuts that enter the construction of a conservative Hamiltonian, while ref.~\cite{Luna:2017dtq} used the BCJ prescription to compute the five-point amplitude relevant to the emission of gravitational radiation (and also presented a method to eliminate the dilaton degree of freedom from the double copy in this approach). More recently, refs.~\cite{Brandhuber:2021kpo,Brandhuber:2021eyq,Brandhuber:2021bsf} presented an algorithmic realisation of the BCJ double copy for heavy mass effective theory, which can be directly applied to the gravitational two-body problem, and which also sheds light on the algebraic structure of the colour-kinematics duality. We encourage the reader to see the gravitational waves section of the companion article on the double copy of amplitudes \cite{SAGEX-22-03}, as well as ref.~\cite{SAGEX-22-13}, which reviews different approaches to describing classical gravity in terms of amplitudes.

\section{Conclusions and future directions}
\label{Conclusions}

As we have seen, the application of scattering amplitudes to classical gravitational physics in recent years has led to significant progress in our
understanding both of observables in classical gravity and of the structure of scattering amplitudes themselves. 
Progress has been rapid, but many questions still remain.
It seems likely that work on these questions will lead to further insights in years to come.

Compact object coalescences are the basic systems of interest in gravitational wave astrophysics. 
These are \emph{bound} systems of black holes or neutron stars. 
In this review we focused instead on scattering processes, because the connection between amplitudes is at its most straightforward in that context.
It is of course possible to link on-shell computations to bound systems, for example using effective theories~\cite{Neill:2013wsa,Cheung:2018wkq,Bern:2019nnu,Bern:2019crd,Bern:2020buy,Bern:2020uwk,Bern:2021dqo,Bern:2021yeh}. 
However a more direct link, presumably building on the analytic continuation highlighted in refs.~\cite{Kalin:2019rwq,Kalin:2019inp,Herrmann:2021lqe,Cho:2021arx}, would be very desirable.

Returning to scattering events, we have seen that the gravitational waveform is an integral of a five-point amplitude. 
This result holds to all perturbative orders; beyond the leading-order example discussed in the text, one needs to subtract cuts of five-point amplitudes.
The connection between the waveform and amplitudes is very welcome as it shows that double-copy and unitarity methods can be used to
determine the waveform. 
One particularly challenging issue for the future is the computation of all the relevant integrals.
The amplitudes themselves are massive five-point amplitudes which involve loop integrals that are extremely challenging.
However, the integrals should simplify in the classical limit~\cite{Herrmann:2021lqe,Herrmann:2021tct}, though it will be important to understand all the relevant classical regions in the loop integrals~\cite{Bern:2021yeh}.
This is not the end of the story: one must integrate over momentum mismatches as for example in equation~\eqref{eq:spectralTensorWaveform}; 
these integrals were clearly elucidated recently in ref.~\cite{Jakobsen:2021smu}.
It would be very interesting to understand the function space resulting from similar integrals at higher orders.

Black holes are not simple pointlike objects. They have event horizons, and can absorb energy and momentum. 
We made no attempt to include these dissipative aspects of black hole physics, 
which nevertheless compete with more traditional post-Minkowskian effects in black hole dynamics~\cite{Goldberger:2019sya,Goldberger:2020wbx,Goldberger:2020geb,Goldberger:2020fot}.
Goldberger and Rothstein have recently developed~\cite{Goldberger:2019sya} an effective field theoretic approach to this issue, 
but it would also be interesting to understand if dissipation can be incorporated more directly in the amplitudes themselves.

Bryce DeWitt famously observed that the only observable in quantum gravity is the $S$ matrix. 
It should therefore be true that \emph{any} observable (which makes sense in both the quantum and the classical theories) 
should determined by scattering amplitudes.
We have seen examples of this: the impulse on point particles, the momentum radiated and of course the waveform are essentially amplitudes. 
The same is true of the change in spin of a particle~\cite{Maybee:2019jus,Bern:2020buy,Kosmopoulos:2021zoq} (and the change in colour in Yang--Mills theories~\cite{delaCruz:2020bbn,delaCruz:2021gjp}).
Can we then recover all of the physical content of the Einstein equation from amplitudes in asymptotically locally Minkowskian spacetimes?

At cosmological scales, the universe appears to be in a de Sitter phase, which presents a fundamental challenge to the amplitudes programme.
There are tentative signs that some on-shell simplifications persist beyond the locally Minkowski case. 
For example, Kerr--Schild metrics play an important role in the classical double copy, and there is no problem generalising Kerr--Schild solutions
to de Sitter space~\cite{Luna:2015paa}. 
Indeed, it is trivial to generalise the Kerr--Schild form of the double copy from Coulomb to Schwarzschild to the de Sitter case.
We have also learned that the double copy, at the level of scattering amplitudes, can be formulated for some classes of non-trivial exact solutions~\cite{Adamo:2017nia,Adamo:2018mpq,Adamo:2021dfg}.
Amplitudes themselves appear as residues in the study of cosmological correlators~\cite{Arkani-Hamed:2018kmz}.
Can we find a fully non-perturbative formulation of the double copy? 

Even in Minkowski space, there are important outstanding puzzles regarding amplitudes, the double copy, and exact solutions.
We saw that the double copy naturally leads to a relation between the leading-order Maxwell and Weyl spinors in on-shell momentum space.
This relationship endures upon computing the integrals for a special class of solutions, including Schwarzschild and Kerr.
Classically, these spacetimes are known to admit Kerr--Schild coordinates, and in these coordinates the double copy appears very simply
as an \emph{exact} relationship of the form $g_{\mu\nu} = \eta_{\mu\nu} + \phi k_\mu k_\nu$, $A_{\mu} = \phi k_\mu$~\cite{Monteiro:2014cda}. 
But from the perspective of scattering amplitudes, it is difficult to understand the role of Kerr--Schild coordinates: indeed, amplitudes
compute gauge-invariant quantities, so there seems to be no role for coordinates. 
Similarly, we know classically that the Weyl spinors for Schwarzschild, Kerr, etc.~are linear in the mass given an 
appropriate choice of frame.
As this linearised Weyl spinor is precisely computed by the quantum theory this suggests that the double copy is an exact relationship
in these cases. How can we understand this exactness from the perspective of the scattering amplitudes?

\section*{Acknowledgments}

This work  was supported  by the European Union's Horizon 2020 research and innovation programme under the Marie Sk\l{}odowska-Curie grant agreement No.~764850 {\it ``{SAGEX}''}.
DOC is supported by the U.K. Science and Technology Facilities 
Council (STFC) grant ST/P000630/1.
DAK's work was supported in part by the French 
\textit{Agence Nationale pour la Recherche\/}, under 
grant ANR--17--CE31--0001--01,
and in part by the European Research Council, under
grant ERC--AdG--885414. RM is supported by a Royal Society University Research Fellowship.

\end{document}